\documentclass{jfm}

\usepackage{graphicx}
\usepackage{amsmath,amsfonts}
\usepackage{natbib}

\renewcommand{\vec}[1]{\ensuremath{\boldsymbol{#1}}}

\newcommand{\rmi}{\ensuremath{\mathrm{i}}}
\newcommand{\rmd}{\ensuremath{\mathrm{d}}}

\newcommand{\aver}[1]{\ensuremath{\Big\langle{#1}\Big\rangle}}

\def\pd{\partial}
\def\ie{{\it i.e., }}
\def\eg{{\it e.g., }}
\def\viz{{\it viz., }}

\def\nb{{\it n.b., }}
\def\cf{{\it cf., }}
\def\la{\Big\langle\,}
\def\ra{\,\Big\rangle}

\ifCUPmtlplainloaded \else
  \checkfont{eurm10}
  \iffontfound
    \IfFileExists{upmath.sty}
      {\typeout{^^JFound AMS Euler Roman fonts on the system,
                   using the 'upmath' package.^^J}%
       \usepackage{upmath}}
      {\typeout{^^JFound AMS Euler Roman fonts on the system, but you
                   dont seem to have the}%
       \typeout{'upmath' package installed. JFM.cls can take advantage
                 of these fonts,^^Jif you use 'upmath' package.^^J}%
      }
  \else
  \fi
\fi

\ifCUPmtlplainloaded \else
  \checkfont{msam10}
  \iffontfound
    \IfFileExists{amssymb.sty}
      {\typeout{^^JFound AMS Symbol fonts on the system, using the
                'amssymb' package.^^J}%
       \usepackage{amssymb}%
       \let\le=\leqslant  
         
      }{}
  \fi
\fi

\ifCUPmtlplainloaded \else
  \IfFileExists{amsbsy.sty}
    {\typeout{^^JFound the 'amsbsy' package on the system, using it.^^J}%
     \usepackage{amsbsy}}
    {\providecommand\boldsymbol[1]{\mbox{\boldmath $##1$}}}
\fi

\newsavebox{\astrutbox}
\sbox{\astrutbox}{\rule[-5pt]{0pt}{20pt}}

\newcommand\tti{\ensuremath{\rightarrow\infty}}

\title{The Elemental Shear Dynamo}

\author[J. C. McWilliams]%
{J\ls A\ls M\ls E\ls S\ns C.\ns Mc\ls W\ls I\ls L\ls L\ls I\ls A\ls M\ls S$^1$%
  \thanks{Email address for correspondence: jcm@atmos.ucla.edu}}

\affiliation{$^1$Department of Atmospheric and Oceanic Sciences, University of California, Los Angeles, CA 90095-1565, USA}

\pubyear{2010}
\volume{650}
\pagerange{119--126}
\date{?; revised ?; accepted ?. - To be entered by editorial office}

\begin{document}

\maketitle

\begin{abstract}
  A quasi-linear theory is presented for how randomly forced,
  barotropic velocity fluctuations cause an exponentially-growing,
  large-scale (mean) magnetic dynamo in the presence of a uniform
  shear flow, $\vec{U} = S x \vec{e}_y$.  It is a ``kinematic'' theory
  for the growth of the mean magnetic energy from a small initial
  seed, neglecting the saturation effects of the Lorentz force.  The
  quasi-linear approximation is most broadly justifiable by its
  correspondence with computational solutions of nonlinear
  magnetohydrodynamics, and it is rigorously derived in the limit of
  large resistivity, $\eta \rightarrow \infty$.  Dynamo action occurs
  even without mean helicity in the forcing or flow, but random
  helicity variance is then essential.  In a sufficiently large domain
  and with small wavenumber $k_z$ in the direction perpendicular to
  the mean shearing plane, a positive exponential growth rate $\gamma$
  can occur for arbitrary values of $\eta$, the viscosity $\nu$, and
  the random-forcing correlation time $t_f$ and phase angle $\theta_f$
  in the shearing plane.  The value of $\gamma$ is independent of the
  domain size. The shear dynamo is ``fast'', with finite $\gamma > 0$
  in the limit of $\eta \rightarrow 0$.  Averaged over the random
  forcing ensemble, the ensemble-mean magnetic field grows more
  slowly, if at all, compared to the r.m.s. field (magnetic energy).
  In the limit of small Reynolds numbers ($\eta, \ \nu \rightarrow
  \infty$), the dynamo behavior is related to the well-known
  alpha--omega {\it ansatz} when the forcing is steady ($t_f
  \rightarrow \infty$) and to the ``incoherent'' alpha--omega {\it
  ansatz} when the forcing is purely fluctuating.
\end{abstract}


\section{Introduction}

This paper presents a theory that yields exponential growth of the
horizontally-averaged magnetic field (\ie a large-scale dynamo) in the
presence of a time-mean horizontal shear flow and a randomly
fluctuating, 3D, barotropic force (\ie with spatial variations only
within the mean shearing plane) in incompressible magnetohydrodynamics
(MHD).  This configuration provides perhaps the simplest paradigm for
a dynamo without special assumptions about the domain geometry or
forcing (\eg without mean kinetic helicity).  We call it the elemental
shear dynamo (ESD).  There is a long history of dynamo theory
\citep{Moffatt78,Krause80,Roberts92,Brandenburg05}, but much of it is
comprised of {\it ad hoc} closure {\it ansatz} (\ie not derived from
fundamental principles and devised for the intended behavior of the
solutions) for how fluctuating velocity and magnetic fields act
through the mean electromotive force curl to amplify the large-scale
magnetic field.  Here the horizontal-mean magnetic field equation is
derived within the ``quasi-linear'' dynamical approximations of
randomly forced linear shearing waves and flow-induced magnetic
fluctuations.

In the standard {\it ansatz} \citep{Moffatt78} , the mean-field
equation in dynamo theory has the functional form of
\begin{equation}
  \pd_t \overline{\vec{B}} = \mathsfbi{L}  \cdot \overline{\vec{B}}  
	+  \mathsfbi{D} : \nabla \overline{\vec{B}} + \dots \,,
\label{eq:ansatz}
\end{equation}
where the over-bar indicates some suitably defined average;
$\overline{\vec{B}}$ is the mean magnetic field; and $\mathsfbi{L}$
and $\mathsfbi{D}$ are second- and third-order tensor operators (often
denoted by $\alpha$ and $\beta$) that express the statistical effects
of the velocity field $\vec{v}$ through the curl of the mean
electromotive force, $\overline{\nabla\times(\vec{v}\times\vec{B})}$.
The dots encompass possible higher-order derivatives of
$\overline{\vec{B}}$ (which would be relatively small if there were a
spatial scale separation between the mean field and the fluctuations)
and resistive diffusion.  If $\vec{v}$ itself is steady in time, then
(\ref{eq:ansatz}) is an exact form for the electromotive effect, and
the kinematic dynamo problem can be viewed as an eigenvalue problem
for the exponential growth rate $\gamma$ given $\overline{\vec{v}}$;
in this case, however, there will be no scale separation between
$\overline{\vec{v}}$ and $\overline{\vec{B}}$, and $\gamma$ may not be
positive.  An important weakness in such an {\it ansatz} is the lack
of justification for particular forms of $\mathsfbi{L}$ and
$\mathsfbi{D}$ in time-dependent flows.  We will see that the ESD
theory provides a clear justification, and it mostly does not fit
within the {\it ansatz} (\ref{eq:ansatz}) because the tensors are
time-integral operators except in particular limits (Sec.
\ref{sec:limit}).

The ESD problem specifies a steady flow with uniform shear $S$, a
small initial seed amplitude and vertical wavenumber $k_z$ for the mean
magnetic field, and a particular horizontal wavenumber $\vec{k}_{\perp
f}$ and correlation time $t_f$ for the random force.  It defines an
ensemble of random-force time series that each gives rise to a
statistically stationary velocity field, and the induced dynamo
behavior is assessed over long integration times with further ensemble
averaging.

This paper takes a general parametric view of the ESD derivation and
solutions. A parallel report utilizing a minimal proof-of-concept
derivation for the treble limit of small kinetic and magnetic Reynolds
numbers and weak mean shear is in \citet{Heinemann11a}; the relation
between the two papers is described in Sec.  \ref{sec:limit.S}.  The
experimental basis for developing the ESD theory is the 3D MHD
simulations in \citet{Yousef08b,Yousef08a}.  They show a large-scale
dynamo in a uniform shear flow with a random, small-scale force at
intermediate kinetic and magnetic Reynolds numbers.  Their dynamo
growth rate is not affected by a background rotation, even Keplerian.
Additionally, new 2$^{+}$D simulations --- a barotropic velocity with
spatial variations only within the mean shearing plane $(x,y)$ and a
magnetic field with $(x,y)$ variations plus a single wavenumber $k_z$
in the vertical direction $z$ perpendicular to the plane --- also
manifests a large-scale dynamo \citep{Heinemann11b}.  Furthermore,
within this 2$^{+}$D model, successive levels of truncation of Fourier
modes in the shearing-plane wavenumber demonstrate that its dynamo
behavior persists even into the quasi-linear situation for which the
mean-field theory is derived here.  Thus, the dynamo solutions of the
ESD theory are a valid explanation for computational dynamo behavior
well beyond the asymptotic limit of vanishing magnetic Reynolds
number.

From general MHD for fluctuations in a shear flow (Sec.
\ref{sec:govern}), a quasi-linear model is developed for shearing
waves (Sec. \ref{sec:dynamics}) and for induced magnetic fluctuations
and the horizontal-mean magnetic field evolution equation with dynamo
solutions (Sec.  \ref{sec:induction}).  Analytic expressions for the
dynamo growth rate $\gamma$ are derived in Sec. \ref{sec:limit} for
several parameter limits, and general parameter dependences are
surveyed in Sec.  \ref{sec:general}.  Section \ref{sec:summary}
summarizes the results and anticipates future generalizations and
tests.

\section{Governing Equations}
\label{sec:govern}

The equations of incompressible MHD are the Navier-Stokes equation
for velocity $\vec{v}$,
\begin{equation}
  \partial_t\vec{v} + \vec{v}\cdot\nabla\vec{v} =
-\,\frac{1}{\rho}\nabla p + \vec{B}\cdot\nabla\vec{B} + \nu\nabla^2\vec{v} + \vec{f} \,,
  \label{eq:Navier-Stokes}
\end{equation}
where $\vec{f}$ is a prescribed forcing function, density $\rho$ is
constant, and pressure $p$ is determined by the constraint,
\begin{equation}
  \nabla\cdot\vec{v} = 0 \,,
  \label{eq:incompressible}
\end{equation}
and the magnetic induction equation for $\vec{B}$ (in velocity units),
\begin{equation}
  \partial_t\vec{B} + \vec{v}\cdot\nabla\vec{B} =
  \vec{B}\cdot\nabla\vec{v} + \eta\nabla^2\vec{B} \,,
  \label{eq:induction}
\end{equation}
with 
\begin{equation}
  \nabla\cdot\vec{B} = 0 \,.
  \label{eq:Bincompressible}
\end{equation}

An exact, conservative solution to the above equations is given by an
unmagnetized, uniform shear flow of the form
\begin{equation}
  \vec{v} = S x\vec{e}_y,\quad\vec{B} = 0 \,,
  \label{eq:background-solution}
\end{equation}
where the shear rate $S$ is a constant in space and time and $\vec{e}$
denotes a unit vector.  To study the dynamics of fluctuations on top
of the background shear flow (\ref{eq:background-solution}), we
rewrite the equations of motion in terms of the velocity fluctuations
$\vec{u}$ defined through
\begin{equation}
  \vec{v} = S x\vec{e}_y + \vec{u} \,.
  \label{eq:velocity-fluctuations}
\end{equation}
Assume that the volume average of $\vec{u}$ is zero.  Substituting
(\ref{eq:velocity-fluctuations}) into (\ref{eq:Navier-Stokes}) and
(\ref{eq:induction}) yields
\begin{equation}
  \mathcal{D}\vec{u} + \vec{u}\cdot\nabla\vec{u} + S u_x\vec{e}_y =
  -\nabla p + \vec{B}\cdot\nabla\vec{B} + \nu\nabla^2\vec{u}
  \label{eq:Navier-Stokes-fluctuations}
\end{equation}
and
\begin{equation}
  \mathcal{D}\vec{B} + \vec{u}\cdot\nabla\vec{B} =
  \vec{B}\cdot\nabla\vec{u} + S B_x\vec{e}_y + \eta\nabla^2\vec{B} \,,
  \label{eq:induction-fluctuations}
\end{equation}
where
\begin{equation}
  \mathcal{D} = \partial_t + S x\partial_y \,.
  \label{eq:curly-D}
\end{equation}

The only explicit coordinate dependence in
(\ref{eq:Navier-Stokes-fluctuations}) and
(\ref{eq:induction-fluctuations}) arises through the differential
operator (\ref{eq:curly-D}), which contains the cross-stream
coordinate $x$. This means that we can trade the explicit
$x$-dependence for an explicit time dependence by a transformation to
a shearing-coordinate frame, defined by
\begin{equation}
  x' = x, \quad y' = y - Stx, \quad z' = z, \quad t' = t \,.
  \label{}
\end{equation}
Partial derivatives with respect to primed and unprimed coordinates are
related by
\begin{equation}
  \partial_{x'} = \partial_x + S t\partial_y \,, \quad
  \partial_{y'} = \partial_y \,, \quad
  \partial_{z'} = \partial_z \,, \quad
  \partial_{t'} = \partial_t + S x\partial_y = \mathcal{D} \,,
\end{equation}
which shows that the explicit spatial dependence is indeed eliminated
in the shearing frame. Therefore in shearing coordinates there are
spatially periodic solutions, in particular a Fourier amplitude and
phase factor, expressed alternatively as
\begin{align}
  \chi(x,y,z,t) &= \mathrm{Re}\left\{\, \hat{\chi}(t)\exp\Bigl[\rmi k_x(t)x 
     + \rmi k_y y + \rmi k_z z\Bigr] \,\right\} 
	\nonumber \\
&= \mathrm{Re}\left\{\, \hat{\chi}(t')\exp\Bigl[\rmi k_{x0}x' 
     + \rmi k_{y} y' + \rmi k_z z'\Bigr] \,\right\} \,,
\label{eq:FT1}
\end{align}
where the transverse wavenumber $k_y$ and the spanwise wave number
$k_z$ are constant in both coordinate frames, but the streamwise
wavenumber $k_x$ varies in time according to $k_x(t) = k_{x0} - S k_y
t$.  For an observer in the unprimed (``laboratory'') coordinate
system, a disturbance that varies along the streamwise direction
stretches out as a result of being differentially advected by the
background shear flow; for an observer in the shearing frame the
Fourier phase has fixed wavenumbers $(k_{x0},k_{y0},k_z)$.

\section{Dynamics}
\label{sec:dynamics}

\subsection{Simplifications}

Guided by the experimental demonstrations of the shear dynamo
\citep{Yousef08b,Yousef08a,Heinemann11b}, we make the following
simplifying assumptions:

\begin{enumerate}

\item
The magnetic field strength is sufficiently small so that there is no
back reaction onto the flow. In this so-called kinematic regime, we
drop the Lorentz force.

\item 
The 3D forcing is restricted to two-dimensional spatial variations in
the horizontal $(x,y)$ plane (\ie barotropic flow with $\partial_z\vec{u} = 
\partial_z p = 0$).  (With this assumption it makes no difference whether the
system is rotating around the $\vec{e}_z$ axis or has a stable
density stratification aligned with $\vec{e}_z$. For these dynamical
influences to matter, $\vec{u}$ has to have 3D spatial dependence.)
In this case the dynamics reduce to forced 2D advection-diffusion
equations for the vertical velocity, $u_z$, and the vertical
vorticity, $\omega_z = \vec{e}_z\cdot(\nabla_\perp\times\vec{u}_\perp)$; \viz
\begin{align}
  \mathcal{D}u_z + \vec{u}_\perp\cdot\nabla_\perp u_z &=
  \nu\nabla_\perp^2 u_z + f_z 
\nonumber \\
  \mathcal{D}\omega_z + \vec{u}_\perp\cdot\nabla_\perp\omega_z &=
  \nu\nabla_\perp^2 \omega_z 
	+ \vec{e}_z\cdot(\nabla_\perp\times\vec{f}_\perp) \,.
\label{eq:2Ddynamics}
\end{align}
We use a notation for a horizontal vector as
\begin{equation}
  \vec{a}_\perp = a_x\vec{e}_x + a_y\vec{e}_y \,.
\end{equation}
Because $\vec{u}$ has no $z$ dependence, the non-divergence condition
reduces to $\nabla_\perp\cdot\vec{u}_\perp = 0$, and we introduce a
streamfunction $\Phi$ for the horizontal velocity and its associated
vertical vorticity:
\begin{equation}
\vec{u}_\perp = \vec{e}_z \times \nabla_\perp \Phi \,, \qquad
\omega_z = \nabla_\perp^2 \Phi \,.
\label{eq:psi}
\end{equation}

\item
Fluctuation advection is neglected in (\ref{eq:2Ddynamics}), so the vertical 
momentum and vorticity balances are linear.
\begin{align}
  \mathcal{D}u_z &= \nu\nabla^2u_z + f_z
  \nonumber \\
  \mathcal{D}\omega_z &= \nu \nabla_{\perp}^2 \omega_z + \vec{e}_z \cdot
  (\nabla_\perp \times \vec{f}_\perp) \,.
  \label{eq:Navier-Stokes-kinematic}
\end{align}

\end{enumerate}

\subsection{Conservative Shearing Waves}
\label{sec:con-wave}

For linearized conservative dynamics ($\vec{f} = 0$, $\nu = 0$),
(\ref{eq:Navier-Stokes-kinematic}) is
\begin{equation}
\mathcal{D} u_z = \mathcal{D} \omega_z = 0 \,.
\label{eq:conservative}
\end{equation}
The Fourier mode solutions are
\begin{align}
u_z &= \mathrm{Re}\left\{\, \hat u_{z0} \, e^{\rmi \phi} \,\right\} 
	\nonumber \\
\omega_z &= \mathrm{Re}\left\{\, \hat \omega_{z0} 
	e^{\rmi \phi} \,\right\} \,,
\label{eq:Fouriermode}
\end{align}
with a phase function that can be alternatively expressed in
shearing or laboratory coordinates as
\begin{equation}
\phi = k_x' x' + \rmi k_y' y' = k_x(t)x + \rmi k_{y0} y \,.
\label{eq:phi-cons}
\end{equation}
The constants $k_x'= k_{x0}$, $k_y' = k_{y0}$, $\hat u_{z0}$, and
$\hat \omega_{z0}$ are set by the initial conditions, and a tilting
$x$-wavenumber is defined by $k_x(t) = k_{x0} - Sk_{y0} t$. From
(\ref{eq:psi}) the associated horizontal velocity is
\begin{equation}
\vec{u}_\perp = \frac{- \, \vec{e}_z \times \vec{k}_\perp(t)}
	{k_\perp^2(t)} \, \mathrm{Re}\left\{\,\rmi \,
	\hat{\mathcal{\omega}}_{z0} e^{\rmi \phi} \,\right\} \,,
\label{eq:Fourier_perp}
\end{equation}
where $k_\perp^2 = k_x^2 + k_{y0}^2$.  Notice that $\vec{u}_\perp(t)$
grows when $k_x(t)/k_{y0} > 0$ by extracting kinetic energy from the
mean shear (an up-shear phase tilt), and it decays when $k_x(t)/k_{y0}
< 0$ (down-shear).  As $t\rightarrow \infty$, $\vec{u}_\perp
\rightarrow 0$ for any $\vec{k}_0$.  This shearing wave behavior is
sometimes called the Orr effect.

\subsection{Single-Mode Forcing}
\label{sec:force}

In a quasi-linear theory the random fluctuations can be Fourier
decomposed into horizontal wavenumbers, and the resulting velocity and
magnetic fields summed over wavenumber.  It suffices to examine a
single wavenumber forcing to demonstrate the ESD process (\cf
(\ref{eq:superpose})).  When $\vec{f}(x,y)$ is restricted to a single
horizontal wavenumber in the laboratory frame $\vec{k}_{\perp f}$, we
have
\begin{equation}
\vec{f} = \mathrm{Re}\left\{\, \hat{\vec{f}}(t) 
	e^{\rmi \phi_f} \,\right\} \,,
\label{eq:sm_forcing}
\end{equation}
where the Fourier coefficient $\vec{\hat{f}}$ is specified from either
a random process.  The spatial phase of the forcing is fixed in
laboratory coordinates:
\begin{equation}
\phi_f = k_{xf} x + k_{yf} y \,.
\label{eq:phif}
\end{equation}
The non-divergence condition on the Fourier coefficient in
(\ref{eq:sm_forcing}) is $\vec{k}_{\perp f}\cdot\vec{\hat{f}}_\perp = 0$;
hence  we can write
\begin{equation}
  \vec{\hat{f}}_\perp = \hat{f}_\perp\vec{e}_{\perp f} \,, \qquad {\rm with}
\qquad \vec{e}_{\perp f} = \frac{\vec{e}_z\times\vec{k}_{\perp f}}{k_{\perp f}}
\end{equation}
the unit vector perpendicular to the forcing wavevector.  Here
$k_{\perp f} = |\vec{k}_{\perp f}|$.  The forcing coefficient is
thus
\begin{equation} 
\vec{\hat{f}} = \hat{f}_\perp\vec{e}_{\perp f} +
\hat{f}_z\vec{e}_z \,.
\end{equation}
Taking the cross product of $\vec{k}_{\perp f}$ with $\vec{\hat{f}}$ yields
\begin{equation}
  \vec{k}_{\perp f}\times\vec{\hat{f}} =
  k_{\perp f}(\hat{f}_\perp\vec{e}_z - \hat{f}_z\vec{e}_{\perp f}) \,.
\end{equation}
This is used to define two further relations.  The forcing coefficient for
vertical vorticity is
\begin{equation}
\hat{o}_z = \vec{e}_z \cdot \rmi \vec{k}_{\perp f}\times\vec{\hat{f}} 
= \rmi k_{\perp f} \hat{f}_\perp \,.
\end{equation}
The spatially-averaged forcing helicity (defined by $H= \la \vec{f}
\cdot \nabla\times\vec{f}\ra^{\vec{x}}$ where brackets denote an
average in the indicated superscript coordinate) associated with a
single Fourier mode is defined by
\begin{equation}
  \hat{H}(t) = \frac{1}{2} \, \mathrm{Re}[ \vec{\hat{f}}^\ast\cdot
  (\rmi\vec{k}_{\perp f}\times\vec{\hat{f}}) ] = \mathrm{Re}[
  \hat{f}_z^\ast \hat{o}_z ] \,,
\label{eq:helicity}
\end{equation}
which is a real number. The asterisk denotes a complex conjugate, and
we now incorporate a caret symbol in $\hat{H}(t)$ to be consistent
with other forcing amplitudes.

The Fourier mode coefficients $\hat{f}_z(t)$ and $\hat{o}_z(t)$ are
complex random time series that are mutually independent between their
real and imaginary parts and between each other, and they have zero
means. We consider an ensemble of many realizations for these time
series.  (We will also analyze solutions with steady forcing (\ie with
$\hat{\vec{f}}$ fixed in time with values taken from the same random
distribution).)  For a given realization, we generate the forcing
coefficients from an Ornstein-Uhlenbeck processes with a finite
correlation time, $t_f$.  Thus,
\begin{eqnarray}
& {\cal E}\Big[ \hat{f}_z^\ast(t_1) \hat{f}_z(t_2) \Big] = F_z \exp\Bigl[
- \, |t_1-t_2|/t_f \Bigr]
	\nonumber \\
& {\cal E}\Big[ \hat{o}_z^\ast(t_1) \hat{o}_z(t_2) \Big] = O_z \exp\Bigl[
- \, |t_1-t_2|/t_f \Bigr]
	\nonumber \\
& {\cal E}\Big[ \hat{f}_z^\ast(t_1) \hat{o}_z(t_2) \Big] = 0 \,,
\label{eq:fcor}
\end{eqnarray}
where ${\cal E}$ is the expectation value averaged over fluctuations
and $F_z$ and $O_z$ are positive forcing variances.  In particular,
the helicity has zero mean, ${\cal E}\Big[ \hat{H}(t) \Big] = 0$.

\subsection{Stochastic, Viscous Shearing Waves}
\label{sec:Stochastic_waves}

We assume single-mode forcing.  For simplicity we assume that the
fluid is at rest at $t=0$.  The resulting solutions to
(\ref{eq:psi})-(\ref{eq:Navier-Stokes-kinematic}) are
\begin{align}
u_z(x,y,t) &= \int_0^t \, \rmd{}\mu \, G_\nu(t,\mu) \, \mathrm{Re}\left\{\,
	\hat{f}_z(\mu) \, e^{\rmi \phi(\mu)} \,\right\}
	\nonumber \\
\omega_z(x,y,t) &= \int_0^t \, \rmd{}\mu \, G_\nu(t,\mu) \, 
\mathrm{Re}\left\{\, \hat{o}_z(\mu) \, e^{\rmi \phi(\mu)} \,\right\} 
	\nonumber \\
\vec{u}_\perp(x,y,t) &= \int_0^t \, \rmd{}\mu \, G_\nu(t,\mu) \,
        \left(\frac{- \, \vec{e}_z \times \vec{k}_\perp(t-\mu)}
	{k_\perp^2(t-\mu)}\right) \, \mathrm{Re}\left\{\,\rmi
	\hat{o}_z(\mu) \, e^{\rmi \phi(\mu)} \,\right\} \,,
\label{eq:velocity}
\end{align}
which can be verified by substitution into the dynamical equations.
The wavevector is $\vec{k}_\perp(t) = (k_x(t),k_{yf})$ with $k_x(t) =
k_{xf} - Sk_{yf} t$ and $k_\perp^2(t) = k_x^2(t) + k_{yf}^2$.  The
phase function $\phi$ represents continuous forcing at the single,
laboratory-frame wavenumber $\vec{k}_{\perp f}$, and its evolving shear
tilting is expressed in $k_x(t)$.  We can write it in either the
sheared or laboratory coordinate frame:
\begin{align}
\phi(x',y',t'; \mu) \ &= \ (k_{xf} + Sk_{yf} \mu)x' + k_{yf}y' 
	\nonumber \\
\phi(x,y,t; \mu) \ \ & = \ k_x(t-\mu) x +  k_{yf}y 
      \ = \ \vec{k}_\perp (t-\mu) \cdot {\bf x}\,,
\label{eq:phase}
\end{align}
where $k_x(t-\mu) = k_{xf} - Sk_{yf} (t-\mu)$.  The viscous damping
effect is expressed by the decay factor,
\begin{equation}
G_\nu(t,\mu) = \exp\Bigl[ - \, \nu \, \int_\mu^t \, \rmd{}\rho
\, k_\perp^2(\rho - \mu)\Bigr] = \exp\Bigl[ - \, \nu \, \int_0^{t-\mu} \, 
\rmd{}\zeta \, k_\perp^2(\zeta)\Bigr]\,,
\label{eq:Enudef}
\end{equation}
which is a Green's function for (\ref{eq:2Ddynamics}).  For
compactness we can write this as an equivalent function of a single
time difference, $G_\nu(t-\mu)$.

In the first line of (\ref{eq:phase}), $\phi$ is expressed in shearing
coordinates $(x',y',t')$; note that the phase of the shearing wave is
independent of $t'$, but it does depend on the forcing at the time
$\mu$ when the wave was spawned.  The second line is the equivalent
expression in laboratory coordinates $(x,y,t)$.  For compactness we
write this below as $\phi(\mu)$, with the other space-time
dependences implicit.

If $\ \nu=0$ (hence $G_\nu=1$) and the forcing is applied only at the
initial instant (\ie $\hat{f}_z = \delta(\mu) \hat{u}_{z0}$ and
$\hat{o}_z = \delta(\mu) \hat{\omega}_{z0}$), then
(\ref{eq:velocity}) reduces to the conservative shearing wave
(\ref{eq:Fouriermode})-(\ref{eq:Fourier_perp}).  For $\nu \ne 0$,
$G_\nu \rightarrow 0$ as $t -\mu \rightarrow \infty$, which implies
the eventual viscous decay of any shearing wave forced at a particular
time $\mu$.

For the dynamo problem we assume that the velocity fluctuations reach
a stationary equilibrium after a finite time, long compared to
$t_f$ and to an approximate viscous decay time, $1/(k_{\perp f}^2
\nu)$.  This formulation implicitly assumes nonzero viscosity, or else
the random velocity variance would grow without limit and not
equilibrate.

\subsection{Kinetic Energy, Non-dimensionalization, and Homogeneity}
\label{sec:KE_ND}

Define the volume-averaged kinetic energy as
\begin{equation}
KE(t) = \frac{1}{2} \, \la \vec{u}^2 \ra^{x,y,z} \,,
\label{eq:KEdef}
\end{equation}
where the angle brackets again indicate an average over the spatial
coordinates.  For this dynamo problem we adopt a dual normalization in
the fluctuation forcing scale and in the resulting velocity scale, or
equivalently the equilibrium kinetic energy:
\begin{equation}
k_{\perp \, f} = 1 \qquad {\rm and} \qquad {\cal E}\Big[ KE \Big] =
\frac{1}{2} \quad {\rm when} \quad t \gg t_f, \ (
	k_{\perp f}^2 \nu)^{-1} \,.
\label{eq:normalize}
\end{equation}
Henceforth, all quantities are made non-dimensional by the implied
length and velocity scales (\ie forcing amplitude, time, magnetic
field amplitude, viscosity, and resistivity).  We further assume, for
definiteness, that the expected value of kinetic energy
(\ref{eq:normalize}) is equally partitioned between the horizontal and
vertical velocity components in (\ref{eq:KEdef}):
\begin{equation}
{\cal E}\Big[ KE_z \Big] = {\cal E}\Big[ KE_\perp \Big] = \frac{1}{4} \,.
\label{eq:part-norm}
\end{equation}
There are no cross-terms in $KE$ because of the statistical
independence of $\hat{f}_z$ and $\hat{o}_z$ in (\ref{eq:fcor}).  This
partition thus gives separate normalization conditions for $F_z$ and
$O_z$.  We will see in Sec. \ref{sec:induction} that both $F_z$ and
$O_z$ must be nonzero for the shear dynamo to exist.

For the solutions in (\ref{eq:velocity}), the kinetic energy density
involves products of Fourier factors, with product phases $\pm
\phi(\mu) \pm \phi(\mu')$, inside a double time-history integral over
$\mu$ and $\mu'$.  The $z$ average is trivially 1 for a barotropic flow
with no $z$ dependence in $\phi$.  We assume the horizontal domain
size $L$ is large compared to the forcing scale, $1/k_{\perp f}$.  For
the terms with summed phases, the $x$ and/or $y$ averages of
$\pm2(k_{xf}x+k_{yf}y)$ are approximately 0 if $Lk_{\perp f} \gg
1$.  (This could also be assured if $Lk_{yf}/2\pi$ has an integer value as
part of a discretization of the forcing; Sec. \ref{sec:limit.S}.)
Focusing on the remaining terms with differenced phases, we take an
$x$ average over phases $\pm (k_x(t-\mu) - k_x(t-\mu'))x =
\pm Sk_{yf}(\mu-\mu')$.  After performing the $z$ and $y$ averages and 
substituting the forcing covariance functions (\ref{eq:fcor}), the
partitioned normalization conditions from (\ref{eq:part-norm}) are
equivalent to
\begin{align}
& F_z \, \int_0^\infty \, \rmd{}\mu \, \int_0^\infty \, \rmd{}\mu' \ 
G_\nu(t-\mu) \, G_\nu(t-\mu') 
	\nonumber \\
& \qquad \qquad \exp\Bigl[ - \, |\mu - \mu'|/t_f \Bigr]\, 
\aver{\, \exp\Bigl[\rmi S k_{yf} (\mu - \mu')x \Bigr] \,}^x \equiv F_z C_z = 1 
	\nonumber \\
& O_z \, \int_0^t \, \rmd{}\mu \, \int_0^t \, \rmd{}\mu' \ 
G_\nu(t-\mu) \, G_\nu(t-\mu') \ \frac{ {\bf k}_\perp (t - \mu) 
\cdot {\bf k}_\perp (t - \mu') }{k_\perp^2(t-\mu) \, k_\perp^2(t-\mu')}
	\nonumber \\
& \qquad \qquad \exp\Bigl[ - \, |\mu - \mu'|/t_f \Bigr] \,
\aver{\, \exp\Bigl[\rmi S k_{yf} (\mu - \mu')x \Bigr] \,}^x \equiv O_z C_\perp 
= 1 \,,
\label{eq:general_renorm}
\end{align}
which are independent of $t$ as $t \rightarrow \infty$.  This defines
the constants $C_z$ and $C_\perp$ that then determine $F_z$ and $O_z$.
It will simplify the dynamo problem in Sec. \ref{sec:mean-field} to
renormalize the random forcing amplitudes by
\begin{equation}
\hat{f}_z^\dagger = C_z^{1/2} \hat{f}_z \,, \qquad
\hat{o}_z^\dagger = C_\perp^{1/2} \hat{o}_z \,, 
\label{eq:f-renorm}
\end{equation}
whose corresponding expected variances are unity, $F_z^\dagger = C_z
F_z = 1$ and $O_z^\dagger = C_\perp O_\perp = 1$, and the associated
expected energies are $KE_z = F_z^\dagger/4$ and $KE_\perp =
O_z^\dagger/4$.

$C_z$ and $C_\perp$ are continuous, finite (if $\nu > 0$), and
positive functions of $S$, $\nu$, $L$, $t_f$, and the forcing
wavenumber orientation angle $\theta_{f}$, 
\begin{equation}
k_{x f} = \cos\theta_{f} \,, \qquad 
k_{y f} = \sin\theta_{f} \,.
\label{eq:forcingk}
\end{equation}
Note that $0 < \theta_{f} < \pi/2$ is an up-shear tilt when $S > 0$,
while $\pi/2 < \theta_{f} < \pi$ is down-shear.  The extreme values
$\theta_f = 0,\pi$ ($k_{yf} = 0$) are not of interest because there is
no shear-tilting in (\ref{eq:phase}) and no dynamo in Secs.
\ref{sec:induction}-\ref{sec:general}.

We could proceed quite generally in all these parameters, but at the
price of considerable complexity.  Various degrees of simplification
are available in different parameter limits, \eg if the domain is
large (as already partly assumed in $Lk_{\perp f} \gg 1$), $\nu
\rightarrow \infty$, $S \rightarrow 0$, or $t_f \rightarrow 0$.  The
simplifications arise from being able to isolate and integrate over
one or more of the factors in (\ref{eq:general_renorm}) while
approximating the time arguments of the other factors as fixed at the
importantly contributing times insofar as they are varying relatively
slowly.

Among all these parameters, the simplifying limit that seems most
physically general and germane is large $L$, with provisionally finite
values for the other parameters.  For the rest of this section and
Secs. \ref{sec:induction}-\ref{sec:dynamo}, we follow this path, and
in Sec. \ref{sec:limit} some additional and alternative limits are
discussed.  On this path we isolate the spatial average factor in
(\ref{eq:general_renorm}) by the $x$-averaging operation
explicit and integrating over its time argument, $\delta = \mu - \mu'$,
asymptotically over a large interval, while setting $\mu \approx \mu'$
for the other factors (because the spatial average factor is small
everywhere that $\delta$ is not). Thus,
\begin{align}
\int \, \rmd{}\delta \, \aver{\, \exp\Bigl[\rmi S k_{yf} 
\delta x \Bigr] \,}^x 
&\approx \int^{-\infty}_{\infty} \, \rmd{}\delta \, 
\frac{1}{L}\int_{-L/2}^{L/2} \, \rmd{}s \, \exp\Bigl[\rmi Sk_{yf} \delta s \Bigr]
	\nonumber \\
&= \int^{-\infty}_{\infty} \, \rmd{}\delta \, \frac{2}{Sk_{yf} \delta L} 
        \sin\left[\frac{Sk_{yf} \delta L}{2} \right] 
= \frac{2\pi}{S k_{yf} L} \,.
\label{eq:spatial-delta}
\end{align}
The final step on the second line is based on the asymptotic integral
of the sine integral function, $Si$ ({\it mathworld.wolfram.com}).  To
achieve this approximate isolation from the viscous and
forcing-correlation factors, we assume $Lk_{yf} S / \nu, \ Lk_{yf} S
t_f \gg 1$, along with the previous assumption for averaging, $Lk_{yf}
\gg 1$ This is not the distinguished limits of small $S$ or $t_f$ in a
finite domain (Sec. \ref{sec:limit.S}), although when taken
successively following (\ref{eq:spatial-delta}) such limits are well
behaved (Sec. \ref{sec:limit.L}).  The relation
(\ref{eq:spatial-delta}) can equivalently but more compactly be
expressed as
\begin{equation}
\aver{\, \exp\Bigl[\rmi S k_{yf} (\mu - \mu')x \Bigr] \,}^x \approx \ 
	C_L \delta(\mu-\mu') \,,
\label{eq:x-avg}
\end{equation}
with $C_L = 2\pi/SLk_{yf}$. 

Inserting (\ref{eq:x-avg}) into (\ref{eq:general_renorm}) yields
\begin{eqnarray}
& C_z = C_L A_z^2\,, \qquad C_\perp = C_L A_\perp^2 \,,
	\nonumber \\
& A_z^2 = \int_0^\infty \, \rmd{}\rho G_\nu^2 (\rho) \,,
\qquad  A_\perp^2 = \int_0^\infty \, \rmd{}\rho G_\nu^2 (\rho)
	k_{\perp f}^{-2}(\rho) \,.
\label{eq:norm_consts}
\end{eqnarray}

After the normalizations (\ref{eq:normalize})-(\ref{eq:part-norm}) and
the large $L$ approximation yielding (\ref{eq:x-avg}), the
non-dimensional parameters of the ESD model are $S$, $\nu$, $t_f$,
and $\theta_f$, plus other quantities related to $\vec{B}$ defined in
Sec. \ref{sec:induction}.  There is no dependence on $L$.

As an aside we examine the ensemble-mean local velocity variance,
${\cal E}\Big[ {\vec{u}^2} (x,y,z,t) \Big]$, which is different from
the domain-averaged $2{\cal E}\Big[ KE \Big]$.  From (\ref{eq:velocity})
and the covariance properties of the random force (\ref{eq:fcor}), \eg
the vertical velocity variance has the expected value at late time,
\begin{equation}
{\cal E}\Big[ u_z^2 \Big] = 
\int_0^\infty \rmd{}\mu \, \int_0^\infty \rmd{}\mu' \,
G_\nu(t-\mu) \, G_\nu(t-\mu') \, F_z \, \exp\Bigl[ - \, |\mu-\mu'|/t_f \Bigr] 
\, \cos[S k_{yf}x (\mu - \mu')] \,.
\label{eq:uzvar}
\end{equation}
This variance is independent of $t$ because nonzero
viscosity renders $\vec{u}$ stationary.  It is independent of $y$ and
$z$, \ie homogeneous in these coordinates.  But the local variance is
not in general homogeneous in $x$.  In the limit $\nu \tti$, the
integrals can approximately be evaluated (as discussed more fully in
Secs. \ref{sec:limit.L} and \ref{sec:general}) to yield a constant
value equal to $F_z^\dagger = F_z C_z$ in (\ref{eq:general_renorm}).
For finite viscosity the peak variance is at $x=0$, and it decreases
with $|x|$ on a scale $\sim \, 1/(Sk_{yf}t_f)$; this can be seen by
taking the limit of small $t_f$ where
\begin{equation}
{\cal E}\Big[ u_z^2 \Big] \approx \frac{2 F_b t_f}{1+ (St_fk_{yf}x)^2} \
\int_0^\infty \rmd{}\mu G_\nu^2(t-\mu) \,.
\end{equation}
Homogeneity is thus restored for small $S$ or small $t_f$, although
these limits are formally incompatible with the approximation
underlying (\ref{eq:x-avg}), which is therefore to be understood as a
horizontal average over a region that encompasses any variance
inhomogeneity.  The fundamental source of forced shearing-wave
inhomogeneity is the special zero value of the mean flow $Sx\vec{e}_y$
at $x=0$: the phase-tilting rate $Sk_{yf}x$ increases with $|x|$,
while the forcing correlation time $t_f$ does not depend on $x$.
Homogeneity holds for $\nu \tti$ because the forced shearing waves
have non-trivial amplitude only for $\phi = \phi_f$, \ie no phase tilting.

An amelioration of the inhomogeneity magnitude results from the
dynamical freedom to add a random forcing phase $r(\mu)$ to
(\ref{eq:phase}); \eg a model for $r$ is a 2$\pi$-periodic random walk
with correlation time $t_r$.  Inhomogeneity is eliminated if $t_r
\rightarrow 0$, but it still occurs with finite $t_r$.  A broader
posing of the ESD problem is for a family of mean flows with the same
mean shear, \ie $\vec{V} = U_*\vec{e}_x + (V_*+S(x-x_*))\vec{e}_y$,
and a corresponding modification of the forced shearing-wave phase
(\ref{eq:phase}) to $\phi(x,y,t;\mu) = k_x(t-\mu) (x - x_*) + k_{yf}
(y-y_*) - \vec{k}_{\perp f} \cdot \vec{V}_*(t-\mu) + 0.5 S U_*
(t-\mu)^2 + r(\mu)$.  An expanded-ensemble average over $\vec{V}$, and
over $x_*$ in particular, restores homogeneity in $x$ of ${\cal
E}\Big[ {\vec{u}^2} \Big]$ for general parameters, which thus is a
corollary of translational and Galilean invariances.  These
generalizations in $r$ and $\vec{V}$ do not change the dynamo behavior
in anything except the shearing-wave phase, which does not appear in
$KE$ or the ESD (Sec. \ref{sec:mean-field} {\it et seq.}), so we now
drop further consideration of them.

\section{Magnetic Induction}
\label{sec:induction}

Write the induction equation (\ref{eq:induction-fluctuations}) as
\begin{equation}
  \mathcal{D}\vec{B} = \nabla\times(\vec{u}\times\vec{B})
  + S B_x\vec{e}_y + \eta\nabla^2\vec{B} \,.
\label{eq:3D-induct}
\end{equation}
To simplify matters, we note that the induction equation is linear in
the magnetic field. Therefore, for a barotropic velocity field $\vec{u}(x,y)$,
the electromotive force does not give rise to any mode
coupling in $z$.  We pose the dynamo problem as exponential growth of
the horizontally-averaged (\ie mean) horizontal magnetic field with an
initial seed amplitude and a single $z$-wavenumber $k_z$,
\begin{equation}
\aver{\vec{B}_\perp}^{x,y}= \mathrm{Re}\left\{\vec{\mathcal{B}}(t)
\, e^{\rmi k_z z} \right\} \,.
\label{eq:averB}
\end{equation}
Thus, both $k_z$ and the initial mean field, $\vec{\mathcal{B}}(0)$,
are parameters of the problem; without loss of generality, we can take
$|\vec{\mathcal{B}} (0) | = 1$ as the non-dimensional normalization of
$\vec{B}$.  Because we are interested in dynamo behavior with
exponential growth, this normalization choice does not affect the
resulting growth rate.  We then define $\theta_B$ as its initial
orientation angle:
\begin{equation}
\mathcal{B}_x (0) = \cos\theta_{B} \,, \qquad \mathcal{B}_y (0) =
\sin\theta_{B} \,.
\end{equation}
Because $\hat{\vec{f}}(t)$ is a stochastic variable, the more precisely
stated dynamo problem is exponential growth of mean magnetic energy
$|\vec{\mathcal{B}}|^2(t)$ looking over many realizations and/or long
time intervals.

Because there is no Fourier mode coupling in $z$, we can assume the entire
magnetic field has only a single $k_z$, and the application of the
gradient operator is simplified to
\begin{equation}
  {\nabla} = \nabla_\perp + \rmi k_z\vec{e}_z \,.
\end{equation}
We only need to solve for the horizontal component of $\vec{B}$, \ie
$\vec{B}_\perp$, and obtain $B_z$ diagnostically from the
solenoidality condition,
\begin{equation}
  B_z = -\, \frac{\nabla_\perp\cdot\vec{B}_\perp}{\rmi k_z} \,.
\label{eq:vertB}
\end{equation}
For the mean field $\aver{\vec{B}_\perp}^{x,y}$, there is no
associated vertical component.  The horizontal induction
equation from (\ref{eq:3D-induct}) is
\begin{equation}
  \mathcal{D}\vec{B}_\perp
  = - (\vec{u}\cdot{\nabla}) \, \vec{B}_\perp +
  (\vec{B}_\perp\cdot\nabla_\perp)\, \vec{u}_\perp
  + SB_x\vec{e}_y + \eta{\nabla}^2\vec{B}_\perp \,.
\label{eq:induction-2D-p}
\end{equation}
Because it is enough to focus on the horizontal components of
$\vec{B}$, we henceforth drop the subscript $\perp$ and interpret all
vectors $\vec{a}$ as horizontal unless indicated otherwise by a
subscript: a 3D vector will be $\vec{a}_3$ (\eg $\nabla_3$).

The non-dimensional parameters in the ESD associated with the magnetic
field are $k_z$, $\eta$, and $\theta_B$; these are in addition to the
dynamic parameters listed at the end of Sec. \ref{sec:KE_ND}.

\subsection{Magnetic Fluctuations}
\label{sec:magfluc}

Decompose the horizontal magnetic field into fluctuation and mean
components,
\begin{equation}
\vec{B}(x,y,z,t) = \vec{\delta B}(x,y,z,t) + 
\mathrm{Re}\left\{\vec{\mathcal{B}}(t)\,e^{\rmi k_z z}\right\} \,.
\label{eq:ESDtruncation}
\end{equation}
For consistency with (\ref{eq:averB}), we specify that
$\aver{\vec{\delta B}}^{x,y} = 0$.  We evaluate the vertical companion
field $\delta B_z$ by (\ref{eq:vertB}).  Because
(\ref{eq:induction-2D-p}) is linear in $\vec{B}$, we see that
$\vec{\delta B}$ will have the same vertical phase factor as the mean
field; \ie we define its complex coefficient $\vec{b}$ by
\begin{equation}
\vec{\delta B} = \mathrm{Re}\left\{\vec{b}(x,y,t) \, 
    e^{\rmi k_z z}\right\} \,.
\label{bkz}
\end{equation}

By assumption the ESD contains only a single phase component for the
horizontal magnetic fluctuation field $\vec{b}(x,y,z,t)$ determined
from the horizontal forcing wave number $\vec{k}_f$ (through its
shear-tilting phase $\phi$ in (\ref{eq:phase})) and the vertical
wavenumber $k_z$ of the seed mean magnetic field.  Its induction
equation from (\ref{eq:induction-2D-p}) is forced by the stochastic
shearing waves and the horizontal mean magnetic field, \ie
\begin{equation}
\mathcal{D} \vec{\delta B} = \vec{\delta F} + S\vec{e}_y \delta B_x
+ \eta {\nabla}_3^2 \vec{\delta B} \,,
\label{eq:b1}
\end{equation}
where the curl of the fluctuation electromotive force $\vec{\delta F}$ is
\begin{align}
\vec{\delta F}(x,y,z,t) &= - u_z \pd_z \aver{\vec{B}}^{x,y} 
   + \left( \aver{\vec{B}}^{x,y} \cdot \nabla \right) \, \vec{u}
	\nonumber  \\
&= -\, u_z \mathrm{Re}\left\{\rmi k_z \vec{\mathcal{B}}
  e^{\rmi k_z z}\right\} 
+ \left( \mathrm{Re}\left\{\vec{\mathcal{B}} e^{\rmi k_z z}\right\}
  \cdot \nabla \right) \vec{u} \,.
\label{eq:Fbdef}
\end{align}
There is no contribution from $-\,  \left( \vec{u} \cdot \nabla \right)
\, \aver{\vec{B}}^{x,y}$ because $\aver{\vec{B}}^{x,y}$ has no
horizontal gradient.  One can view the ESD fluctuation induction
equation (\ref{eq:b1}) for $\vec{\delta B}$ as a first-iteration
approximation to the full MHD induction in the presence of $\vec{u}$
and $\aver{\vec{B}}^{x,y}$; \ie it is a projection of MHD onto a
magnetic field with only the shearing-wave phase and a horizontally
uniform component.

This simplified equation for $\vec{b}$ is the heart of the
quasi-linear ESD theory (\ie linear for magnetic fluctuations,
nonlinear for the horizontal mean).  The quasi-linear simplification
can be rigorously justified only if $|\vec{b}| \ll
|\vec{\mathcal{B}}|$, in which case all higher harmonics of the phases
in $\vec{b}$ will be negligibly small compared to the primary phase;
this condition is met in the limit $\eta \tti$, \ie vanishing magnetic
Reynolds number (Sec.  \ref{sec:limit}).  In the next section we will
see how the spatially-averaged induction from the shearing waves
induces dynamo growth in $\vec{\mathcal{B}}$.  This quasi-linear
theory is formally incomplete when the preceding justification
condition is not always well satisfied by its solutions.
Nevertheless, they correspond to the shear dynamo found in 2$^+$D and
3D simulations for a fairly broad range of parameters
\citep{Yousef08b,Yousef08a,Heinemann11b}, so we infer that the ESD
provides a cogent explanation of the dynamo process even beyond its
rigorously derivable limit.  When $\vec{u}$ variance is inhomogeneous
(Sec. \ref{sec:KE_ND}), $\vec{\delta B}$ variance will be so as well.

Using the shearing wave solution (\ref{eq:velocity}) and the mean
field expression in (\ref{eq:ESDtruncation}) and an analogous vertical
phase factor decomposition for $\vec{\delta F}$ as for $\vec{\delta
  B}$ in (\ref{bkz}), we evaluate the fluctuation forcing term as
\begin{align}
\vec{F}_b (x,y,t) &= 
\int^t_0 \, \rmd{}\mu \, G_\nu(t-\mu) \,
\Bigl[\  - \, \rmi k_z \vec{\mathcal{B}}(t) \mathrm{Re}\left\{ \hat{f}_z(\mu) 
	 \, e^{\rmi \phi(\mu)} \,\right\} 
\nonumber \\
& + \, \frac{\vec{e}_z \times \vec{k}(t-\mu)}{k^2(t-\mu)} \, 
       (\vec{k}(t-\mu) \cdot \vec{\mathcal{B}}(t)) \, 
	\mathrm{Re}\left\{\, \hat{o}_z(\mu) \, 
	\, e^{\rmi \phi(\mu)} \, \right\} \ \Bigr] \,.
\label{eq:Fbdef-2}
\end{align}
{\it Pro tem} we do not yet use the renormalized forcings
(\ref{eq:f-renorm}) but will do so in the next section.  The two
right-side lines here are, respectively, from the two terms in the
second line of (\ref{eq:Fbdef}).  The magnetic fluctuation Fourier
phases are thus $\pm \, \phi(\mu)+ k_z z$ where $\phi$ is the shearing
wave phase in (\ref{eq:phase}).

We can write the solution of (\ref{eq:b1}) for $\vec{b}$
analytically.  Again utilizing the vertical phase factorization
(\ref{bkz}), we have
\begin{align}
\vec{b}(x,y,z,t) &= 
	\int^t_0 \, \rmd{}\lambda \, \int^\lambda_0 \rmd{}\mu \
	G_\eta(t-\mu,\lambda-\mu)G_\nu(\lambda-\mu)
\nonumber \\
& 
\Bigl[\ - \, \rmi k_z \mathsfbi{S}(t-\lambda) \cdot \vec{\mathcal{B}}(\lambda)
      \, \mathrm{Re}\left\{\, \hat{f}_z(\mu) \, e^{\rmi \phi(\mu)} \,\right\} 
\nonumber \\
+ \, \frac{\vec{e}_z \times \vec{k}(t-\mu)}{k^2(\lambda-\mu)} &\, 
     (\, \vec{k}(\lambda-\mu) \cdot \vec{\mathcal{B}}(\lambda) \,) \,
   \mathrm{Re}\left\{\, \hat{o}_z(\mu) \, e^{\rmi \phi(\mu)}  
   \,\right\} \ \Bigr] \,.
\label{eq:mag-fluc-2}
\end{align} 
Here we define the second-order, real tensor $\mathsfbi{S}$ by
\begin{equation}
\mathsfbi{S} (t) = \mathsfbi{I} + S t \vec{e}_y \vec{e}_x \,,
\label{Stensor}
\end{equation}
with $\mathsfbi{I}$ the identify tensor, and the resistive decay
factor (another Green's function) by
\begin{equation}
G_\eta(t,\lambda,\mu) =  \exp\Bigl[ - \, \eta \, \int_\lambda^t \, \rmd{}\rho
\, k_3^2(\rho - \mu)\Bigr]  =  \exp\Bigl[ - \, \eta \, 
\int_{\lambda-\mu}^{t-\mu} \, \rmd{}\zeta \, k_3^2(\zeta)\Bigr] 
\label{eq:Eeta}
\end{equation}
with $k_3^2(t) = k^2(t) + k_z^2$.  Again, for compactness we write
this as $G_\eta(t-\mu,\lambda-\mu)$.  Thus, in the quasi-linear ESD,
$\vec{\delta B}$ is an induced magnetic shearing wave arising from
$\vec{u}$ and $\aver{\vec{B}}^{x,y}$.

\subsection{Mean Field Equation}
\label{sec:mean-field}

The governing equation is the horizontal average of
(\ref{eq:induction-2D-p}):
\begin{equation} 
\partial_t \aver{\, \vec{B} \,}^{x,y} = \aver{\, \vec{F}_{B} \,}^{x,y}
  + S \aver{\, B_x \,}^{x,y} \vec{e}_y 
  - \eta k_z^2 \aver{\, \vec{B} \,}^{x,y} \,,
\label{eq:meanB}
\end{equation}
where
\begin{equation}
\aver{\, \vec{F}_{B} \,}^{x,y}(z,t) = \aver{\, 
  - \, (\vec{u} \cdot{\nabla})\vec{b}^\prime - \,
 (u_z \partial_z )\vec{b} ^\prime
  + \, (\vec{b}^\prime\cdot\nabla)\vec{u} \,}^{x,y} \,.
\label{eq:FBdef}
\end{equation}
Because of the horizontal average in the ESD mean-field equation,
there is no representation of any spatial structure associated with
wave-averaged inhomogeneity in the electromotive force curl
(Secs. \ref{sec:KE_ND} and \ref{sec:magfluc}).

The induction forcing itself depends linearly on $\vec{\mathcal{B}}$
through $\vec{b}$ in (\ref{eq:mag-fluc-2}), where it enters in a
time-history integral.  So (\ref{eq:meanB}) is a linear
integral-differential equation for $\vec{\mathcal{B}}(t)$, for which
no general analytic solution is known.  Instead, we evaluate the
expression for $\vec{F}_\mathcal{B}$ below and obtain a double-time
integral, second-order tensor operator on $\vec{\mathcal{B}}(t)$ that
we will solve numerically in general (Sec. \ref{sec:dynamo}) and
analytically in certain limits (Sec. \ref{sec:limit}).  This yields
a closed-form equation for the mean magnetic field amplitude as a
function only of the forcing time histories, $\hat{f}_z(t)$ and
$\hat{o}_z(t)$, and the parameters $\vec{k}_{f}$, $S$, $\eta$,
and $\nu$.

As with the $\vec{b}$ solution in the preceding section, the
derivation for $\aver{\, \vec{F}_\mathcal{B} \,}^{x,y}$ is rather
elaborate.  It involves substituting the shearing wave solution
(\ref{eq:velocity}) and the magnetic fluctuation (\ref{eq:mag-fluc-2})
into (\ref{eq:FBdef}) and performing the horizontal average by
identifying the zero horizontal phase components and applying
(\ref{eq:x-avg}); these details are in Appendix \ref{sec:appA}.  If we
again define a vertical Fourier coefficient $\vec{F}_\mathcal{B}$, as
in (\ref{eq:averB}), the result is
\begin{align}
  \vec{F}_\mathcal{B}(z,t) &= - \, \frac{C_L}{2} \, \int^t_0 \,
  \rmd{}\lambda \, \int^\lambda_0 \rmd{}\mu \
  G_\eta(t-\mu,\lambda-\mu)\, G_\nu(\lambda-\mu) \, G_\nu(t-\mu)
  \nonumber \\
  & \quad \Bigl[\ |\hat{f}_z|^2(\mu)\, k_z^2 \, 
   \mathsfbi{S}(t-\lambda) \cdot \vec{\mathcal{B}}(\lambda) 
\ + \,
  \rmi k_z \, \mathrm{Re}\left\{\hat{f}_z^\ast(\mu)\hat{o}_z(\mu)\right\}
  \, \vec{e}_z \times \vec{k}(t-\mu)
  \nonumber \\
  & \qquad \quad \Bigl(\,
 \frac{\vec{k}(\lambda-\mu)}{k^2(\lambda-\mu)} \cdot \vec{\mathcal{B}}(\lambda)
\, + \, \frac{\vec{k}(t-\mu)}{k^2(t-\mu)} \cdot
  \mathsfbi{S}(t-\lambda) \cdot
    \vec{\mathcal{B}}(\lambda)
  \,\Bigr) \ \Bigr] \,.
\label{eq:FBevaluation}
\end{align}
Notice that the forcing helicity $\hat{H}(\mu)$ from
(\ref{eq:helicity}) plays a prominent role.  With the solutions in
Secs. \ref{sec:limit}-\ref{sec:general}, we find there is only
transient algebraic growth in $\vec{\mathcal{B}}(t)$ (\ie no dynamo)
when the forcing helicity is zero.  Therefore, there is no dynamo if
either $\hat{f}_z$ or $\hat{o}_z$ is zero.  In fact, the induced
magnetic fluctuations from a horizontal velocity field, forced by
$\hat{o}_z$ only, have no effect at all on $\vec{\mathcal{B}}$.

Now simplify $\vec{F}_\mathcal{B}$ and the $\mathcal{B}$ equation
by the forcing renormalization (\ref{eq:f-renorm}) augmented by
the following related quantities:
\begin{equation}
{\cal F}^\dagger = \frac{1}{2} \, |\hat{f}_z^\dagger|^2 \,, \qquad
{\cal H}^\dagger = \frac{A_z}{2A_\perp} \, 
\mathrm{Re}\left\{\hat{f}_z^{\dagger \ast}(\mu)\hat{o}_z^\dagger (\mu)\right\}
= \frac{C_z}{2} \hat{H} \,, \qquad G_\nu^\dagger = \frac{1}{A_z^2} G_\nu \,.
\label{eq:FHG_renorm}
\end{equation}
With these the mean electromotive force curl becomes
\begin{align}
  \vec{F}_\mathcal{B}(z,t) &= - \, \int^t_0 \,
  \rmd{}\lambda \, \int^\lambda_0 \rmd{}\mu \
  G_\eta(t-\mu,\lambda-\mu)\, G_\nu^\dagger(\lambda-\mu)\, G_\nu^\dagger(t-\mu)
  \nonumber \\
  & \quad \Bigl\{\ {\cal F}^\dagger(\mu)\, k_z^2 \, 
   \mathsfbi{S}(t-\lambda) \cdot \vec{\mathcal{B}}(\lambda) 
\ + \,
  \rmi k_z \, {\cal H}^\dagger(\mu) \, \vec{e}_z \times \vec{k}(t-\mu)
  \nonumber \\
  & \qquad \quad 
\Bigl[\, k^{-2}(\lambda-\mu) + k^{-2}(t-\mu) \,\Bigr]
\Bigl(\, \vec{k}(\lambda-\mu) \cdot \vec{\mathcal{B}}(\lambda)\,\Bigr) 
\ \Bigr\} \,.
\label{eq:FBevaluation2}
\end{align}
An identity used for the final term is $\vec{k}(t-\mu) \cdot 
\mathsfbi{S}(t-\lambda) \cdot \vec{\mathcal{B}}(\lambda) \, = \,
\vec{k}(\lambda-\mu) \cdot \vec{\mathcal{B}}(\lambda)$.

After factoring the structure $\mathrm{Re}\left\{ \ \cdot \
e^{\rmi k_z z}\right\}$ from (\ref{eq:meanB}), the
equation for the complex amplitude $\vec{\mathcal{B}}(t)$ becomes
\begin{align}
& \partial_t \vec{\mathcal{B}} = S \mathcal{B}_x \vec{e}_y 
	- \eta k_z^2 \vec{\mathcal{B}}
	\nonumber \\
- & \, 
	\int^t_0 \, \rmd{}\lambda \, \int^\lambda_0 \rmd{}\mu \
G_\eta(t-\mu,\lambda-\mu)G_\nu^\dagger(\lambda-\mu)G_\nu^\dagger(t-\mu) \,
\Bigl\{\, {\cal F}^\dagger(\mu)\, k_z^2  \mathsfbi{S}(t-\lambda) \cdot 
\vec{\mathcal{B}}(\lambda) \ +
	\nonumber \\
& \,
\rmi k_z {\cal H}^\dagger(\mu)  \, \vec{e}_z \times 
\vec{k}(t-\mu) \,
\Bigl[\, k^{-2}(\lambda-\mu) + k^{-2}(t-\mu) \,\Bigr]
\Bigl(\, \vec{k}(\lambda-\mu) \cdot \vec{\mathcal{B}}(\lambda)\,\Bigr) 
\, \Bigr\} \,.
\label{eq:complexB}
\end{align}
A final compaction step is to factor out the resistivity effect
associated with the vertical wavenumber by defining
\begin{equation}
\vec{\mathcal{B}}(t) = \widetilde{\vec{\mathcal{B}}}e^{-\eta k_z^2 t} \,.
\label{eq:factor-eta}
\end{equation}
This modifies (\ref{eq:complexB}) to
\begin{align}
& \partial_t \widetilde{\vec{\mathcal{B}}} 
	= S \widetilde{\mathcal{B}}_x \vec{e}_y 
	\nonumber \\
- & \, 
	\int^t_0 \, \rmd{}\lambda \, \int^\lambda_0 \rmd{}\mu \
	\widetilde{G}_\eta(t-\mu,\lambda-\mu)
	G_\nu^\dagger(\lambda-\mu)G_\nu^\dagger(t-\mu) \,
\Bigl\{\, {\cal F}^\dagger(\mu)\, k_z^2  \mathsfbi{S}(t-\lambda) \cdot 
     \widetilde{\vec{\mathcal{B}}}(\lambda) \ +
	\nonumber \\
& \,
\rmi k_z {\cal H}^\dagger(\mu)  \, \vec{e}_z \times 
	\vec{k}(t-\mu) \,
\Bigl(\, k^{-2}(\lambda-\mu) + k^{-2}(t-\mu) \,\Bigr)
\Bigl(\, \vec{k}(\lambda-\mu) \cdot 
	\widetilde{\vec{\mathcal{B}}}(\lambda)\,\Bigr) 
\, \Bigr\} \, ,
\label{eq:complexB2}
\end{align}
where $\widetilde{G}_\eta$ is the resistive decay associated with the
horizontal wavevector, defined analogously to $G_\eta$ with
$k_3(\zeta)$ replaced by $k(\zeta)$ in (\ref{eq:Eeta}), \ie factoring
out the decay associated with $k_z$, 
\begin{equation}
G_\eta(t,\lambda,\mu) = \exp\Bigl[- \, \eta k_z^2 (t - \lambda)\Bigr]
\, \widetilde{G}_\eta(t,\lambda,\mu) \,.
\label{eq:Gtilde}
\end{equation}

The functional form of (\ref{eq:complexB2}) is
\begin{equation}
\pd_t \widetilde{\vec{\mathcal{B}}} 
= \mathsfbi{L}  \cdot \widetilde{\vec{\mathcal{B}}}(t) 
	+ \int_0^t \, \rmd{}\lambda \ \mathsfbi{J}(t,\lambda) \cdot
	\widetilde{\vec{\mathcal{B}}}(\lambda) \,,
\label{eq:ansatz2}
\end{equation}
where $\mathsfbi{L}$ and $\mathsfbi{J}$ are second-order tensors.  This ESD
form differs from the common {\it ansatz} (\ref{eq:ansatz}) by the
time-history integral, but it does fit within the formal framework
analyzed by \citet{Sridhar09} for velocity fields whose
dynamical origin was unspecified (in contrast to our particular case
of shearing wave velocities).  We show in Sec. \ref{sec:limit} that
the common {\it ansatz} is recovered in our ESD theory in the limit of
$\eta, \nu \rightarrow \infty$.  The definitions of the $\mathsfbi{L}$ and
$\mathsfbi{J}$ tensors are
\begin{align}
\mathsfbi{L}_{mn} &= S\, \delta_{my}\delta_{nx}
	\nonumber \\
\mathsfbi{J}_{mn}(t,\lambda) &= - \, \int^\lambda_0 \rmd{}\mu \
	\widetilde{G}_\eta(t-\mu,\lambda-\mu)
		G_\nu^\dagger(\lambda-\mu)G_\nu^\dagger(t-\mu) \,
	\nonumber \\
& \Bigl[\, {\cal F}^\dagger(\mu)\, k_z^2  \, 
	\left(\, \mathsfbi{S}_{mn}(t-\lambda \,\right)
      + \, \rmi k_z {\cal H}^\dagger(\mu)  \,
	\nonumber \\
& \qquad \quad 	\Bigl(\, k^{-2}(\lambda-\mu) + k^{-2}(t-\mu) \,\Bigr) \,
	k_\ell(t-\mu) \, k_n(t-\lambda) \,\epsilon_{z\ell m} \,\Bigr]
\label{eq:LI}
\end{align}
for horizontal indices, $\{ m,n,\ell \} = \{ x,y \}$, and the usual
Kronecker delta and Levi-Civita epsilon tensors.  $\mathsfbi{L}$ contains
the background shear effect on $\widetilde{\vec{\mathcal{B}}}$,
while $\mathsfbi{J}$ contains the mean electromotive force resulting from
the random barotropic forces and induced magnetic fluctuations.
$\mathsfbi{S}_{mn} = \delta_{mn} + S(t-\lambda)\delta_{my}\delta_{nx}$ is
as defined in (\ref{Stensor}).

The ESD (\ref{eq:complexB}) is invariant with respect to several sign
symmetries in the forcing, wavenumber, initial conditions, and mean
shear.  Because the random forcing amplitudes, $\hat{f}_z(t)$ and
$\hat{o}_z$, are statistically symmetric in sign, a change of sign in
either one implies ${\cal H}^\dagger \leftrightarrow - \, {\cal
H}^\dagger$, ${\cal F}^\dagger \leftrightarrow {\cal F}^\dagger$, and
the statistical distribution of $\vec{\mathcal{B}}(t)$ will be
unchanged.  In addition there are the following invariances for
particular realizations of the ESD: (i) $(k_z,\ {\cal H}^\dagger)
\leftrightarrow - \, (k_z, \ {\cal H}^\dagger)$; (ii) $\vec{k}_f
\leftrightarrow - \, \vec{k}_f$; (iii) $\vec{\mathcal{B}}
\leftrightarrow - \, \vec{\mathcal{B}}$; and (iv) $(S, \ {\cal H}^\dagger, \
k_{xf}, \ \mathcal{B}_x) \leftrightarrow - \, (S, \ {\cal H}^\dagger,
\ k_{xf}, \ \mathcal{B}_x)$ with $(k_{yf}, \ \mathcal{B}_y)
\leftrightarrow (k_{yf}, \ \mathcal{B}_y)$.

Because the ESD is a quasi-linear theory based on Fourier
orthogonality in $k_z$ and $\vec{k}_f$, it satisfies a superposition
principle; the full MHD equations
(\ref{eq:Navier-Stokes})-(\ref{eq:Bincompressible}) do not allow
superposition, of course.  The functional form of the superposition is
a generalization of (\ref{eq:averB}) and (\ref{eq:ansatz2}):
\begin{align}
\aver{\vec{B}_\perp}^{x,y}(z,t) &= \sum_{k_z} \, 
      \mathrm{Re}\left\{\widetilde{\vec{\mathcal{B}}}(k_z,t)
       \exp\Bigl[-\eta k_z^2 t + \rmi k_z z\Bigr]\right\} \,,
    \nonumber \\
\pd_t \widetilde{\vec{\mathcal{B}}}(k_z,t) &= 
          \mathsfbi{L}(k_z)  \cdot \widetilde{\vec{\mathcal{B}}}(k_z,t) 
	+ \int_0^t \, \rmd{}\lambda \ 
    \left(\, \sum_{\vec{k}_f} \mathsfbi{J}(k_z,\vec{k}_f,t,\lambda) \,\right) \cdot
	\widetilde{\vec{\mathcal{B}}}(k_z, \lambda) \,.
\label{eq:superpose}
\end{align}
The random force $\hat{\vec{f}}(\vec{k}_f,t)$ in (\ref{eq:sm_forcing}) is
assumed to be statistically independent for each $\vec{k}_f$
component with whatever normalization is chosen in place of the
single-component normalization (\ref{eq:normalize}).

\subsection{Dynamo Behavior}
\label{sec:dynamo}

A numerical code has been written to solve the ESD in
(\ref{eq:complexB2}).  Its algorithm is described in Appendix
\ref{sec:method}.  As expected from the 3D and 2$^+$D full PDE
solutions, a dynamo often occurs when $S$ and ${\cal H}(t)$ are
nonzero.  We now demonstrate a typical dynamo solution, deferring the
more general examination of the ESD parameter dependences until Sec.
\ref{sec:general}, after first obtaining analytic solutions in Sec.
\ref{sec:limit} in certain limiting cases.

\begin{figure}
  \centerline{\includegraphics[width=0.8\textwidth]{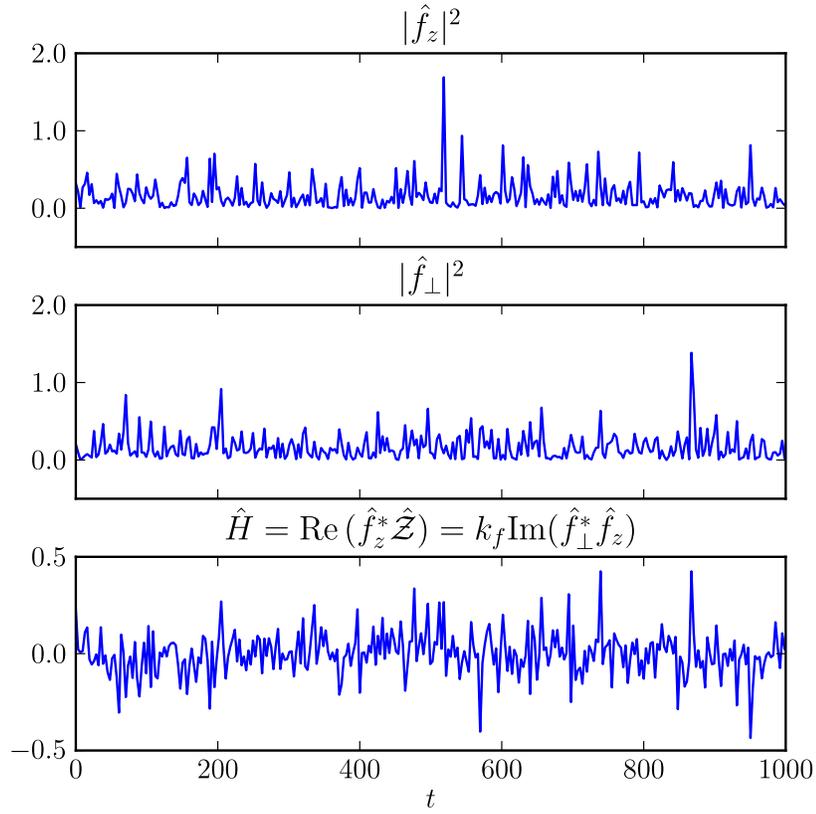}}
  \caption{Time series of the random forcing time variances,
  $|\hat{f}_z|^2$ and $|\hat{f}_\perp|^2$, and forcing
  helicity $\hat{H}(t)$ for a case with $\theta_f=\pi/4$, $S = 1$,
  $t_f= 0.1$, and $\nu=0.1$.  The discrete time step size is
  $\Delta t=0.025$.}
\label{fig:forcing}
\end{figure}

\begin{figure}
  \centerline{\includegraphics[width=0.8\textwidth]{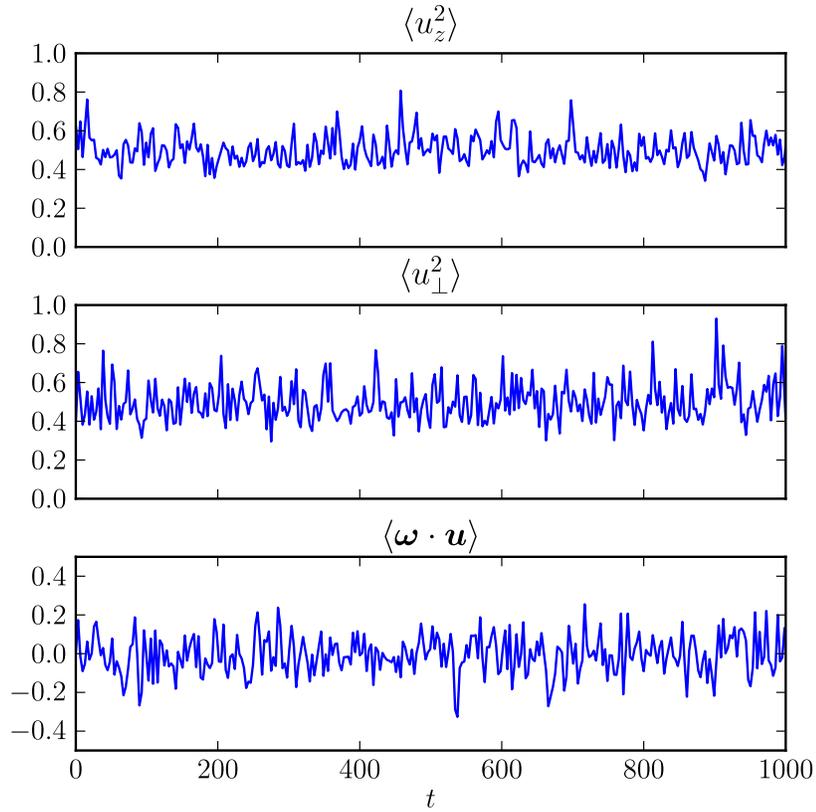}}
  \caption{Time series for vertical and horizontal velocity variances
  (\ie twice $KE_z(t)$ and $KE_\perp(t)$) and the associated kinetic
  helicity response, $\la {\bf u} \cdot {\bf \omega} \ra^{bf x} \,(t)$,
  for random velocities generated by the forcings in Fig.
  \ref{fig:forcing}.}
\label{fig:velocity}
\end{figure}

An illustration of a random realization of the forcings, velocity
variances, and helicity time series is in
Figs. \ref{fig:forcing}-\ref{fig:velocity}.  These are for a case with
moderately up-shear forcing wavenumber orientation ($\theta_f =
\pi/4$), moderately small correlation time $t_f=0.1$ and viscosity
$\nu = 0.1$, and intermediate mean shear rate ($S=1$).  The amplitude
normalizations from (\ref{eq:part-norm}) are evident, as is the
vanishing of the time-averaged helicity.  Because $t_f\nu \ll 1$,
the time scale of the velocity fluctuations is controlled primarily by
the viscous decay time modified by the shear tilting in the $k_x(t)$:
in (\ref{eq:Enudef}) the initial exponential linear decay rate, $\nu =
0.1$, is at first slowed as $k_x$ passes through zero at $t = 1/ S
\tan\theta_f = 1$ and then augmented toward a exponential cubic
decay with a rate coefficient $\approx \, (\nu S^2 k_{yf}^2/3)^{1/3} =
0.26$.

\begin{figure}
  \centerline{\includegraphics[width=0.7\textwidth]{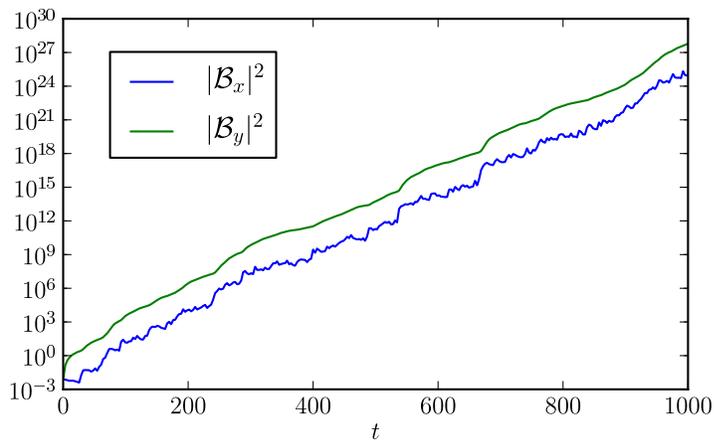}}
  \caption{Mean-field variance time series for the same case as in
  Figs. \ref{fig:forcing}-\ref{fig:velocity}.  Additional case
  parameters are $\eta=0.1$, $k_z = 0.125$, and $\theta_B = \pi/4$.}
  \label{fig:mean-field}
\end{figure}

To obtain a dynamo in (\ref{eq:complexB2}), the vertical wavenumber
$k_z$ must be small but finite; we show below that this is true for
general parameters.  With $k_z = 0.125$ and moderately small $\eta =
0.1$, the time series of the mean magnetic field component variances
are shown in Fig. \ref{fig:mean-field} for the same realization of the
forcing and velocity as in Figs. \ref{fig:forcing}-\ref{fig:velocity}.
There is evident exponential growth in both components of
$\vec{\mathcal{B}}(t)$, \ie this is a dynamo.  If we make an
exponential fit over a long time interval with $|\vec{\mathcal{B}}|
\propto e^\gamma t$, we obtain the same value of $\gamma \approx 0.03$
for each component.  $\vec{\mathcal{B}}(t)$ also manifests a
stochastic variability inherited from the random forcing, and its
fluctuations about the exponential growth exhibit power even at much
lower frequencies than are evident in the forcing and velocity time
series.

The magnitude of $\mathcal{B}_y$ is larger than of $\mathcal{B}_x$ in
Fig. \ref{fig:mean-field}.  This is a common behavior for magnetic
fields in shear flow. A partial and somewhat simplistic explanation is
as a consequence of the first right-side shear term in
(\ref{eq:complexB2}).  A simplified (non-dynamo) system with arbitrary
forcing $\vec{R}(t)$,
\begin{equation}
\partial_t \vec{\mathcal{B}} = S \mathcal{B}_x \vec{e}_y + \vec{r}(t) \,,
\qquad \vec{\mathcal{B}}(0) = \vec{\mathcal{B}}_0 \\,
\end{equation}
has the solution,
\begin{equation}
\vec{\mathcal{B}}(t) = \vec{\mathcal{B}}_0 + \int_0^t \, \rmd{}t' \, 
\vec{R}(t') 
+ S \vec{e}_y \left(\, \mathcal{B}_{x 0} t + \int_0^t \, \rmd{}t' \, 
\int_0^{t'} \, \rmd{}t'' \, r_x(t'') \, \right) \,.
\label{eq:simple}
\end{equation}
The last term $\propto S \vec{e}_y$ will make $|\mathcal{B}_y| \gg
|\mathcal{B}_x$ at late time for most $\vec{R}(t)$.  This anisotropy
effect carries over to the ESD but also involves further right-side
$\vec{\mathcal{B}}$ coupling absent in (\ref{eq:simple}); a coupled
explanation for the anisotropy in dynamo solutions is made in Sec.
\ref{sec:limit.L}.  The initial condition $\vec{\mathcal{B}}_0$ is
usually not dominant in (\ref{eq:simple}) at late time.  The initial
condition is even less important for $\vec{\mathcal{B}}(t)$ in Fig.
\ref{fig:mean-field}, which is obtained with $\theta_B = \pi/4$; in
particular, $\theta_B$ does not determine the dynamo growth rate
$\gamma$.

$\vec{b}(t)$ (not shown) also shows exponential growth in its
amplitude, with $|b_y|$ typically much larger than $|b_x|$ for
the same reason as just explained.  $\vec{b}(t)$ has comparable time
dependence to $u_z(t)$ and $\vec{u}(t)$, as well as an additional
resistive decay influence from $\eta$ and modulations by the
exponential growth and slow variation in $\vec{\mathcal{B}}(t)$.

\section{Dynamo Analysis in Limiting Cases}
\label{sec:limit}

\subsection{$L \rightarrow \infty$;  $\eta, \ \nu \rightarrow \infty$}
\label{sec:limit.L}

The ESD in Secs. \ref{sec:dynamics}-\ref{sec:induction} is based on an
assumption that the horizontal domain size is large, $L \rightarrow
\infty$ (\nb the average of a Fourier exponential in
(\ref{eq:x-avg})).  As a means of obtaining a more readily analyzed
form of the ESD (\ref{eq:complexB2}), we take the additional limit of
$\eta \rightarrow \infty$.  This limit does not change the forcing
amplitude nor the velocity field (Sec.  \ref{sec:dynamics}), which are
independent of $\eta$, but it allows an elimination of one of the time
integrals in the expression for $\vec{b}$ in (\ref{eq:mag-fluc-2}) and
in the equation (\ref{eq:complexB2}) for
$\widetilde{\vec{\mathcal{B}}}$.  It also makes the quasi-linear
approximation rigorously accurate because it yields $|\vec{\delta B}|
\ll |\vec{\mathcal{B}}|$ (as explained after (\ref{eq:Mform})).

The essence of the $\eta \rightarrow \infty$ approximation is that
first the order of integration in (\ref{eq:complexB2}) is reversed,
\[
\int^t_0 \, \rmd{}\lambda \, \int^\lambda_0 \rmd{}\mu = 
\int^t_0 \, \rmd{}\mu \, \int^t_{t-\mu} \, \rmd{}\lambda \,,
\]
and then the $\lambda$ integral is performed by assuming that
$\widetilde{G}_\eta$ is more rapidly varying in $\lambda$ than any of the
other integrand factors and furthermore is nonzero only when $\lambda
\rightarrow t$, \ie $t-\lambda = O(\eta^{-1})$.  We evaluate this
approximation as
\begin{equation}
\int^t_{t-\mu} \, \rmd{}\lambda \, \widetilde{G}_\eta(t-\mu,\lambda-\mu)
\rightarrow \frac{1}{\eta k^2(t-\mu)}
\end{equation}
for all $\mu \ne t$ (the integral is zero for $\mu = t$) and set the
$\lambda$ arguments of other factors in the integrand to $t$.  With
this approximation, the $(L,\eta)$-limiting form of
(\ref{eq:complexB2}) becomes
\begin{align}
& \partial_t \widetilde{\vec{\mathcal{B}}} = S \widetilde{\mathcal{B}}_x(t) 
\vec{e}_y - \, \frac{1}{\eta} \, \int^t_0 \, \rmd{}\mu \ 
	\frac{G_\nu^{\dagger 2}(t-\mu)}{k^2(t-\mu)} \,
	\nonumber \\
& \qquad
\Bigl[\, k_z^2  {\cal F}^\dagger(\mu)\, \widetilde{\vec{\mathcal{B}}}(t) \ +
2 \rmi k_z {\cal H}^\dagger(\mu)  \ 
\frac{\vec{e}_z \times \vec{k}(t-\mu)}{k^{2}(t-\mu)}  \
\vec{k}(t-\mu) \cdot \widetilde{\vec{\mathcal{B}}}(t) \,\Bigr] \,.
\label{eq:complexB-eta}
\end{align}
This is a purely differential equation for $\widetilde{\vec{\mathcal{B}}}(t)$; 
\ie it matches the common {\it ansatz} form in (\ref{eq:ansatz}), \viz 
\begin{equation}
\pd_t \widetilde{\vec{\mathcal{B}}} = 
\mathsfbi{L} \cdot \widetilde{\vec{\mathcal{B}}}(t) \,,
\label{eq:Mform}
\end{equation}
for the identifiable single-time, second-order tensor $\mathsfbi{L}(t)$
that contains a time-history integral in $\mu$ over the random
forcing.

An analogous simplification of the expression for $\vec{b}$ in
(\ref{eq:mag-fluc-2}) can be made, with the result that $\vec{b}
\propto 1/\eta$.  This gives the important analytic result that the
quasi-linear approximation to (\ref{eq:induction}) is asymptotically
convergent as $\eta \rightarrow \infty$; the higher harmonics of the
shearing-wave Fourier phase ($\pm m \phi$, $m > 1$) generated in
$\vec{b}$ by the fluctuation electromotive term are $O(\eta^{-m})$,
hence negligible compared to the mean-field term proportional to
$\aver{\vec{B}}^{x,y}$ in (\ref{eq:Fbdef}).

\begin{figure}
  \centerline{\includegraphics[width=0.7\textwidth]{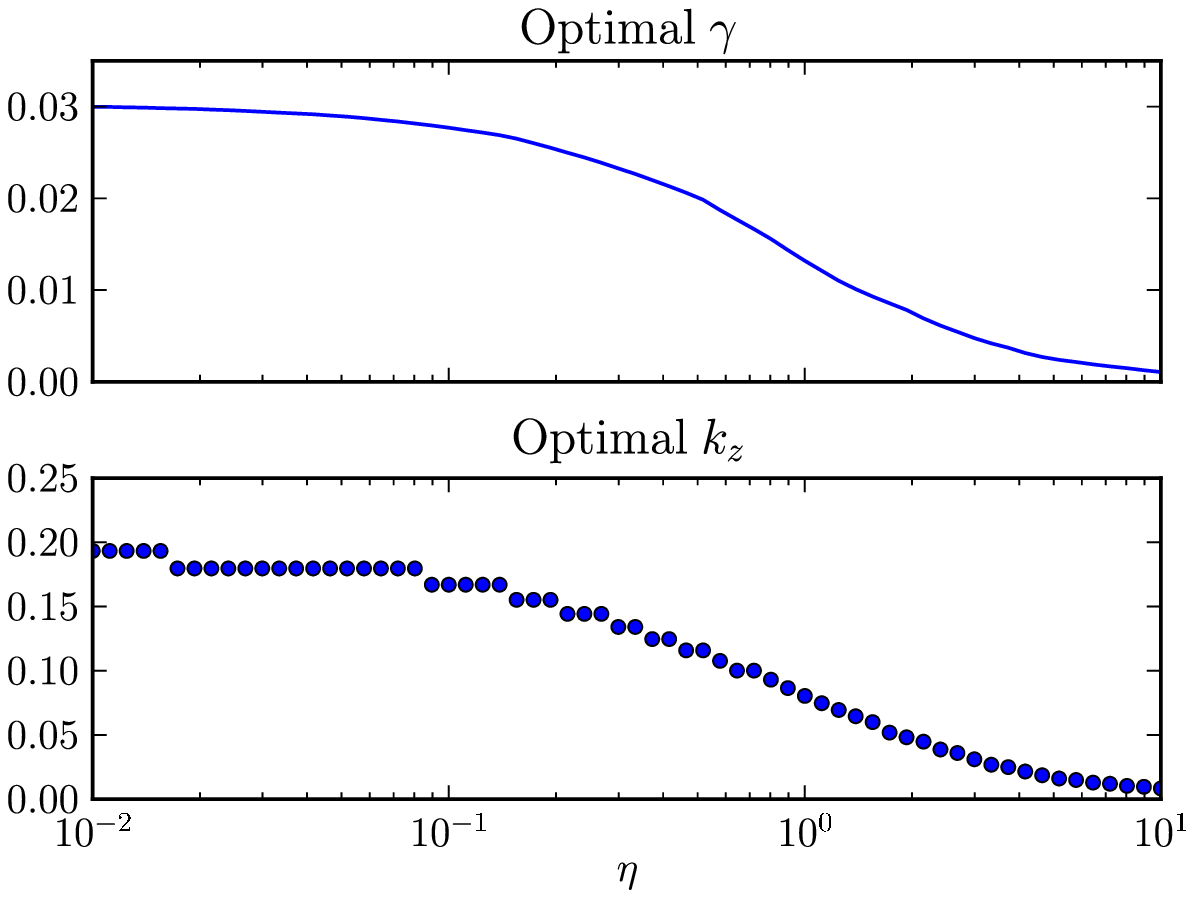}}
  \caption{Mean-field growth rate $\gamma$ associated with the optimal
    $k_z$ value as a function of resistivity $\eta$.  Other parameters
    are $\theta_f=\pi/4$, $S=1$, $t_f=0.1$, $\eta=0.1$, and $\Delta
    t=0.025$.  }
  \label{fig:gamma-eta}
\end{figure}

Numerical solutions of (\ref{eq:complexB-eta}) exhibit dynamo behavior
similar to the example in Sec. \ref{sec:dynamo}, and the parameter
dependences for $\gamma$ are similar to those described in Sec.
\ref{sec:general} for the general ESD.  In particular, $\gamma$ is
small here because $\eta$ is large, in contrast to the ``fast dynamo''
limit where $\gamma$ becomes independent of $\eta$ (\cf
Fig. \ref{fig:gamma-eta}).

To obtain further analytic simplicity we can take a sequential limit
of (\ref{eq:complexB-eta}) as $\nu \rightarrow \infty$.  As with the
$\eta$ limit, this selects an integration time $\mu \approx t$, where
the viscous decay factor is integrated out by the approximate relation
for large $t$,
\begin{equation}
\int^t_0 \, \rmd{}\mu \, G_\nu^2(t-\mu) \rightarrow \frac{1}{2\nu} \quad 
{\rm or} \quad 
\int^t_0 \, \rmd{}\mu \, G_\nu^{\dagger 2}(t-\mu) \rightarrow 1 \,,
\end{equation}
utilizing the renormalization relations in (\ref{eq:norm_consts}) and
(\ref{eq:FHG_renorm}).  The $(L,\eta,\nu)$-limit mean-field equation
from (\ref{eq:complexB-eta}) is
\begin{align}
\partial_t \widetilde{\vec{\mathcal{B}}} &= 
S \widetilde{\mathcal{B}}_x(t) \vec{e}_y  - \, \frac{1}{\eta} \, 
\Bigl[\, k_z^2  {\cal F}^\dagger(t)\, \widetilde{\vec{\mathcal{B}}}(t) \ +
2 \rmi k_z {\cal H}^\dagger(t)  \  (\vec{e}_z \times \vec{k}_f)
\  \vec{k}_f \cdot \widetilde{\vec{\mathcal{B}}}(t) \, \Bigr] \,,
\label{eq:complexB-eta-nu}
\end{align}
after using $k^2(0) = k_f^2 = 1$ from (\ref{eq:normalize}).  In the
tensor representation (\ref{eq:Mform}), $\mathsfbi{L}(t)$ is defined
for (\ref{eq:complexB-eta-nu}) by
\begin{equation}
\mathsfbi{L} = S \, 
	    \begin{pmatrix} 0 & 0 \cr
             1 &  0 \cr \end{pmatrix}
	\, - \, \frac{k_z^2  {\cal F}^\dagger(t)}{\eta} \, 
             \begin{pmatrix} 1 & 0 \cr
              0 & 1 \cr \end{pmatrix}
	\, - \, \frac{2 \rmi k_z {\cal H}^\dagger(t)}{\eta}
   \begin{pmatrix} \cos\theta_f \sin\theta_f & \sin^2\theta_f \cr
 -\, \cos^2\theta_f & - \, \cos\theta_f \sin\theta_f \cr \end{pmatrix} \,,
\label{eq:L-B-t}
\end{equation}
after a substitution for $\vec{k}_f$ from (\ref{eq:forcingk}).  All of
the forcing time history in the coefficient tensor $\mathsfbi{L}(t)$
has now disappeared.  The history integral also disappears in the
companion $\vec{b}$ formula derived from (\ref{eq:mag-fluc-2}).
Furthermore, there is no remaining dependence on $\nu$ in
(\ref{eq:complexB-eta-nu}) because ${\cal F}^\dagger$ and ${\cal
H}^\dagger$ are $O(1)$ quantities by the $KE$ normalization in
(\ref{eq:normalize}) and the forcing renormalization in
(\ref{eq:f-renorm}) and (\ref{eq:FHG_renorm}).  Large $\eta$ and $\nu$
values lead to momentum and induction equation balances with
negligible time tendency terms and negligible shear tilting in
$\vec{k}(t)$ because $\phi \rightarrow \phi_f$ and ${\bf k}(t)
\rightarrow {\bf k}_f$.

We now consider two further limits in the forcing correlation time
$t_f$ that yield analytic expressions for $\gamma$.

\subsubsection{Steady Forcing} 
\label{sec:steady}

\begin{figure}
  \centerline{\includegraphics[width=0.7\textwidth]{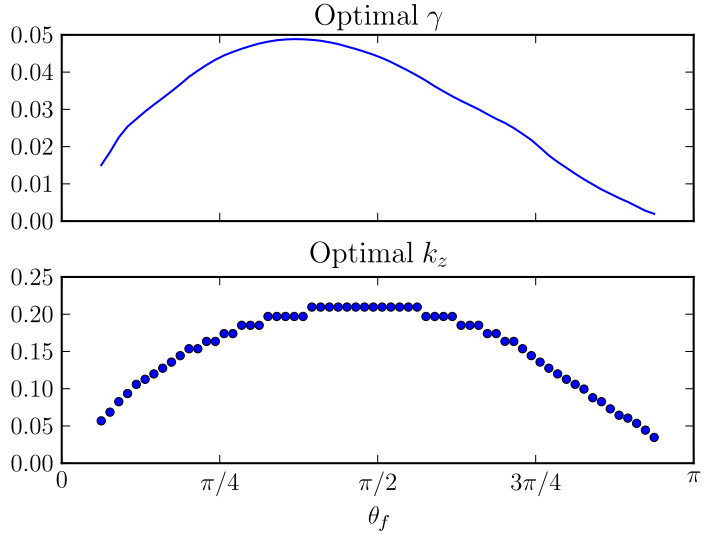}}
  \caption{Mean-field growth rate $\gamma$ associated with the optimal
    $k_z$ value as a function of forcing angle $\theta_f$. Other
    parameters are $\theta_f=\pi/4$, $S=1$, $t_f=0.1$, $\nu=1.0$,
    and $\Delta t=0.025$.  }
  \label{fig:gamma-theta}
\end{figure}

Suppose the forcing values taken from the random distributions in
Sec. \ref{sec:force} but are held steady in time; this is a limit
based on the physical approximation that the forcing amplitudes change
more slowly than the inverse growth rate for the dynamo, $\gamma t_f
\gg 1$.  In this limit (\ref{eq:complexB-eta-nu})-(\ref{eq:L-B-t}) has
its $\mathsfbi{L}$ independent of time, hence there are eigensolutions
with
\begin{equation}
\widetilde{\vec{\mathcal{B}}} \ \propto \ e^{\Gamma t } \,.
\end{equation}
The eigenvalues of $\mathsfbi{L}$ are
\begin{equation}
\Gamma = - \, \frac{k_z^2}{\eta} {\cal F}^\dagger  \ \pm \
\left( \frac{2 \rmi k_z \sin^2\theta_f {\cal H}^\dagger \, 
S }{\eta} \right)^{1/2} \,.
\label{eq:Mev}
\end{equation}
The dynamo growth rate for total mean field $\vec{\mathcal{B}}$ is
defined as the largest real part of $\Gamma$ plus a correction of
$- \, \eta k_z^2$ from the transformation in (\ref{eq:factor-eta}):
\begin{equation}
\gamma = - \left( \eta + \frac{{\cal F}^\dagger}{\eta} \right) \, k_z^2  + 
\left(  \frac{k_z | S {\cal H}^\dagger |  \sin^2\theta_f }{\eta}
\right)^{1/2} \,.
\label{eq:steady}
\end{equation}
The first term is negative and the second positive.  A dynamo occurs
with $\gamma > 0$ if there are both forcing helicity and shear and if
$k_z$ is small enough but nonzero.  With $S=0$, there is no
dynamo. For $|S|$ above a critical-shear threshold value,
\begin{equation}
S_{cr} = \frac{\eta k_z^3}{\sin^2\theta_f} \, \frac{(\eta + \eta^{-1}
{\cal F}^\dagger)^2}{|{\cal H}^\dagger|} > 0 \,,
\end{equation}
$\gamma$ increases with $S$, asymptotically as $\sqrt{S}$ when the other
parameters are held constant, and $\gamma$ decreases with $\eta$ as
$1/\eta$.  For given $S$, there is a lower threshold value for $\eta$
to have a dynamo.  Nonzero forcing helicity is necessary for a dynamo,
but its sign does not matter.  $\gamma = 0$ for $k_z = 0$, and $\gamma
< 0$ for $k_z$ large.  Within an intermediate range where $\gamma >
0$, the optimal $k_z$ and its associated growth rate are
\begin{align}
k_{z \, opt} &= \left( \frac{| S {\cal H}^\dagger|  \sin^2\theta_f}
{16\, \eta (\eta + \eta^{-1} {\cal F}^\dagger)^2}\right)^{1/3}
\approx  \ \left( \frac{| S {\cal H}^\dagger|  \sin^2\theta_f}
	{16\, \eta^3}\right)^{1/3}
	\nonumber \\
\gamma_{opt} &= \left(
\frac{27 \, | S {\cal H}^\dagger|^2 \sin^4[\theta_f]}
{256 \, \eta^2 (\eta + \eta^{-1} {\cal F}^\dagger)} \right)^{1/3}
\approx  \ \left(
\frac{27 \, | S {\cal H}^\dagger|^2 \sin^4[\theta_f]}
{256 \, \eta^3} \right)^{1/3} \,,
\label{eq:steady-opt}
\end{align}
where the approximations are based on neglecting ${\cal F}^\dagger)/\eta^2$.
The optimal $k_z$ decreases with increasing $\eta$.  (In a general MHD
simulation with fixed $(S,\eta,\nu)$ values, all $k_z$ are available,
and the ones supporting a dynamo will emerge in the evolution.)  The
vertical forcing variance ${\cal F}^\dagger$ reduces the dynamo, while
the forcing helicity amplitude $|{\cal H}^\dagger|$ enhances it.
${\cal F}^\dagger$ enters (\ref{eq:L-B-t}) and (\ref{eq:steady})
exactly as an enhanced resistivity; however, the effect is small as
$O(\eta^{-2})$ when ${\cal F}^\dagger = O(1)$ in this large $\eta$
limit.  This is an anisotropic turbulent eddy resistivity acting on
the mean field in the direction perpendicular to the shear plane as a
result of the shearing-wave vertical velocity
\citep{Parker71,Moffatt78}.  The horizontal force $\vec{f}$ acting by
itself has no effect; it makes ${\cal F} = |{\cal H}| = 0$, hence
$\gamma < 0$ (no dynamo).  $\gamma$ is largest where $k_{y f}$ is
largest at $\theta_{f} = \pi/2$; in Sec. \ref{sec:general} we show
that $\gamma$ is usually larger for $\theta_f < \pi/2$
(Fig. \ref{fig:gamma-theta}) because of a dynamo enhancement by the
shear-tilting Orr effect when $\nu < \infty$.  $k_{x f}$ does not
explicitly enter the formula for $\gamma$ in the present case.

The system (\ref{eq:complexB-eta-nu})-(\ref{eq:L-B-t}) in its
steady-helicity limit is a close analog of the so-called alpha--omega
dynamo for galactic disks \citet{Parker71,Kulsrud10}.  Using a mixed
notation from these two sources and assuming a vertical structure
$\aver{\, \vec{B} \,}^{x,y} \ \propto \ e^{\rmi k_z z}$, an ODE
system analogous to (\ref{eq:Mform}) results, with
\begin{equation}
\mathsfbi{L}^{\alpha\Omega} = 
	    \begin{pmatrix} - \, \widetilde{\eta} k_z^2 & \rmi k_z \alpha \cr
             \Omega &  - \, \widetilde{\eta} k_z^2 \cr \end{pmatrix} \,.
\label{eq:Mtensor-aO}
\end{equation}
For constant $\alpha$ and $\Omega$, its eigenvalues are
\begin{equation}
\Gamma^{\alpha\Omega} = - \, k_z^2 \widetilde{\eta} \pm 
\left( \rmi k_z \alpha  \Omega \right)^{1/2} \,.
\label{eq:Mev-aO}
\end{equation}
The correspondence with (\ref{eq:Mev}) is evident with appropriate
identifications between $(\alpha,\ \Omega,\ \widetilde{\eta})$ and
$(\eta^{-1} {\cal H}^\dagger,\ S, \ \eta + \eta^{-1} {\cal
  F}^\dagger)$.  However, the ODE systems are not isomorphic except in
the special case of $k_{xf} = 0$ in (\ref{eq:Mform}).  Thus, in the
steady-forcing ESD, the shear $S$ plays the role of $\Omega$, helical
forcing ${\cal H}^\dagger$ plays the role of $\alpha$, and ${\cal
  F}^\dagger$ plays the role of a turbulent eddy resistivity that
augments the effect of $\eta$. 

The physical paradigm in this paper is random forcing.  Therefore,
even if the forcing is steady in time, it is taken from a random
distribution, and we can ask what the expected value is for
$\widetilde{\vec{\mathcal{B}}}$ (\ie having factored out the resistive
decay in (\ref{eq:factor-eta}), which is not dominant for small
$k_z$).  To answer this we now neglect the turbulent resistivity by
${\cal F}^\dagger$, which is shown above to be a small effect for
large $\eta$.  The eigenvalue (\ref{eq:Mev}) of the tensor
(\ref{eq:L-B-t}) is for a particular forcing value, which we now
generalize to an ensemble distribution,
\begin{equation}
\Gamma(\varepsilon) = \pm \gamma (1 + \rmi s ) \,, \quad
\gamma(\varepsilon) = \frac{1}{\sqrt{2}} \, E S \sin\theta_f > 0 \,, 
\label{eq:Gdistrib}
\end{equation}
with a composite parameter that is a rescaled helicity forcing,
\begin{equation}
\varepsilon = \frac{2k_z{\cal H}^\dagger}{S\eta} \equiv E^2 s \,.
\end{equation}
$E^2$ is the magnitude of $\varepsilon$, and $s = \pm 1$ is its sign.
Consistent with the Ornstein-Uhlenbeck process for the forcing
amplitudes (Sec.  \ref{sec:force}), $\varepsilon$ has a Gaussian
probability distribution function,
\begin{equation}
{\cal P}(\varepsilon) = \frac{1}{\sqrt{2\pi\varepsilon_0^2}} \
\exp\Bigl[ - \, \varepsilon^2/2\varepsilon_0^2 \Bigr] \,, \qquad
\int_{-\infty}^{\infty} \, {\cal P} \,  \rmd{}\varepsilon = 1 \,,
\label{eq:epspdf}
\end{equation}
with an expected variance $\varepsilon_0^2$.  Utilizing ${\cal
  E}\Bigl[{\cal H}^{\dagger 2}\Bigr] = 0.5 \, F_z\dagger \,
O_z^\dagger = 0.5$ from the remark after (\ref{eq:f-renorm}), we
obtain
\begin{equation}
\varepsilon_0^2 = \frac{2k_z^2}{S^2\eta^2} \rightarrow 
\frac{1}{4S^{4/3}\eta^4} \,,
\end{equation}
where the arrow indicates substitution of $k_z^{opt}$ from
(\ref{eq:steady-opt}).  We analyze the dynamo solutions with general
$\varepsilon_0$, but for large $\eta$, $\varepsilon_0$ is expected to
be small.  After a large elapsed time $t_e$, the dynamo solution is
dominated by its leading eigenmode with
$\mathrm{Re}\left\{\,\Gamma\right\} = \gamma > 0$ for any $E \ne 0$.
Neglecting the decaying mode, we write the late-time solution in
vector form as
\begin{equation}
\begin{pmatrix} \widetilde{\mathcal{B}}_x(\varepsilon,t_e) \cr
                 \widetilde{\mathcal{B}}_y (\varepsilon,t_e) \cr \end{pmatrix}
\ = \ {C}_0 \, e^{\gamma t_e} \, \left( \cos[\gamma t_e]
    + i s \sin[\gamma t_e] \right) \
\begin{pmatrix} (1+\rmi s) \gamma/S 
        + \rmi \varepsilon \cos\theta_f \sin\theta_f \cr
                 1 - \rmi \varepsilon \cos^2\theta_f \cr \end{pmatrix} \,.
\label{eq:ef-te}
\end{equation}
${C}_0$ is a complex constant determined from the initial condition,
\begin{equation}
{C}_0 = \frac{1}{\sqrt{2}E(i+\rmi s)} \, 
\left\{\, \widetilde{\mathcal{B}}_x(0) + \widetilde{\mathcal{B}}_y(0) \, 
\left(\frac{(1+\rmi s) \gamma/S - \rmi\varepsilon  \cos\theta_f 
\sin\theta_f}{1 - \rmi \varepsilon \cos^2\theta_f}\right)\,\right\} \,.
\label{eq:C0}
\end{equation}
With (\ref{eq:epspdf}) and (\ref{eq:ef-te}), we can evaluate the
expected value of any property of $\widetilde{\vec{\mathcal{B}}}(t_e)$
and its corresponding distribution $D$ with $\varepsilon$; \eg for the
mean-field vector magnitude,
\begin{equation}
B^{rms} \, \equiv \, {\cal E}\Big[\, 
	|\widetilde{\vec{\mathcal{B}}}|\,(t_e) \, \Big] \, 
= \, \int_{-\infty}^{\infty} \, 
|\widetilde{\vec{\mathcal{B}}}|(\varepsilon,t_e) \, {\cal P}(\varepsilon) 
\,  \rmd{}\varepsilon \equiv \int_{-\infty}^{\infty} \, 
D[\,|\widetilde{\vec{\mathcal{B}}}| \,] \,  \rmd{}\varepsilon \,.
\label{eq:Brms}
\end{equation}

\begin{figure}
  \includegraphics[width=0.49\textwidth]{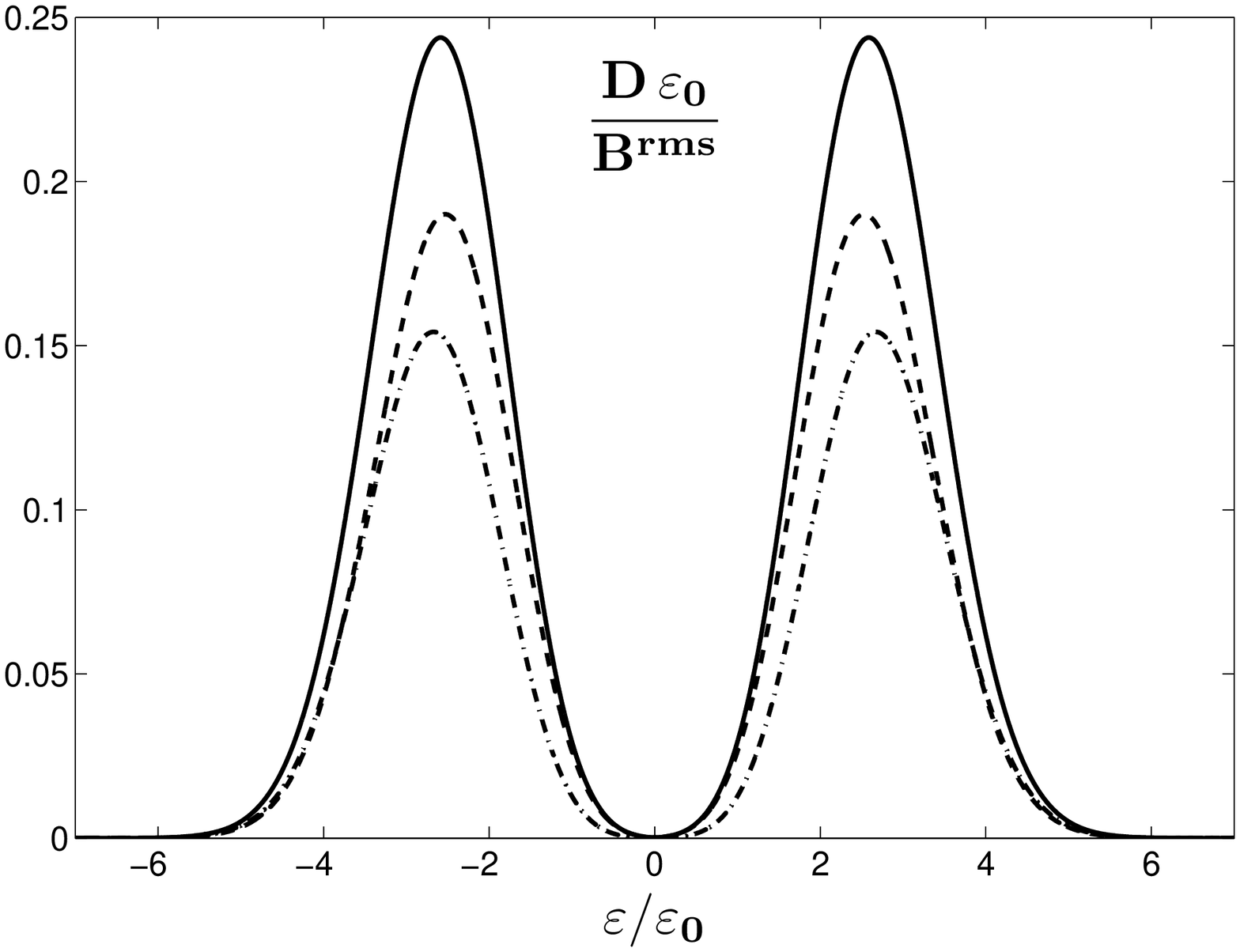}
  \hskip 1pt
  \includegraphics[width=0.49\textwidth]{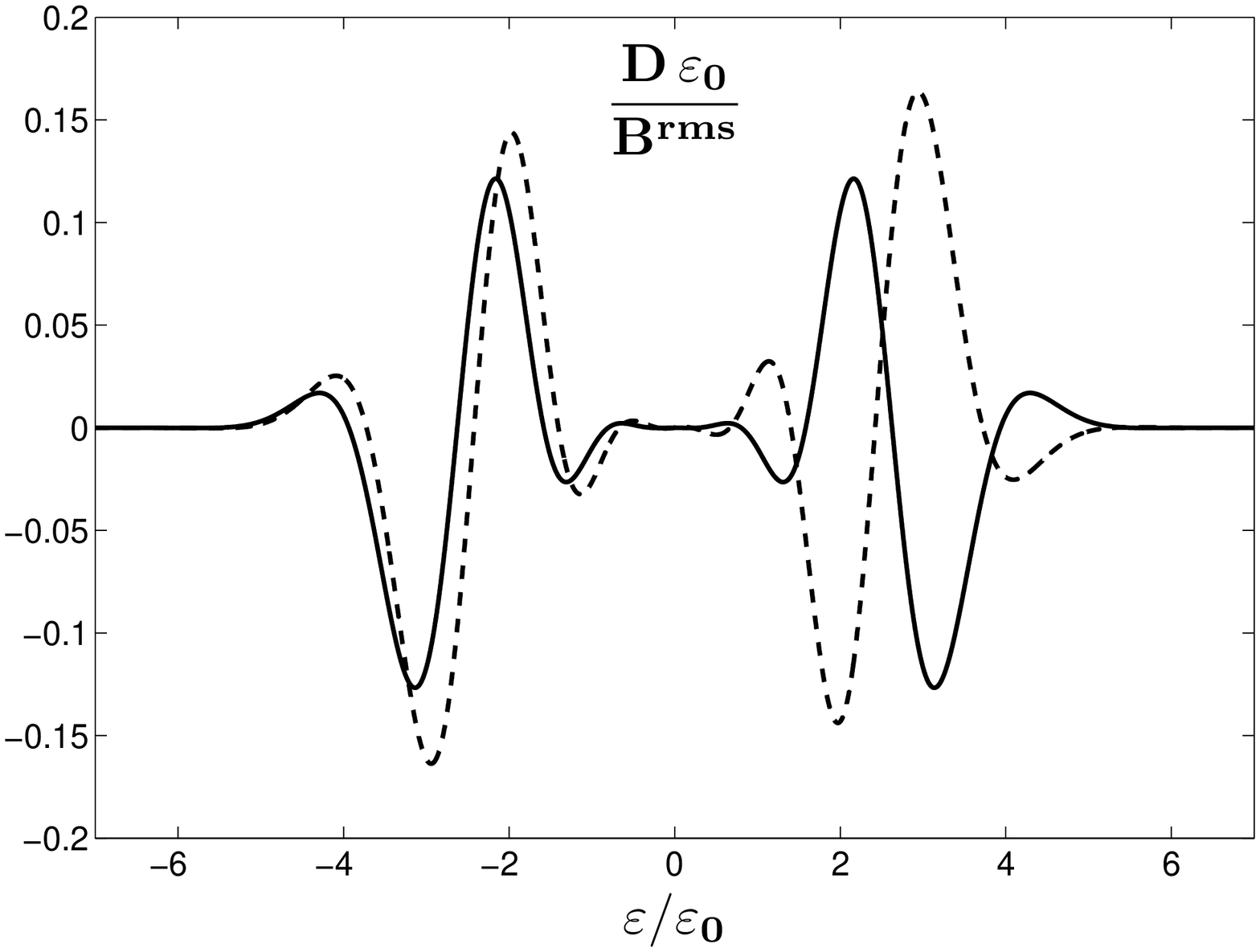}
  \caption{Steady-forcing normalized distributions $D(\varepsilon)$
    for (left) $|\widetilde{\vec{\mathcal{B}}}|$ (solid) and the
    component magnitudes, $|\widetilde{\mathcal{B}}_x|$ (dash-dot) and
    $|\widetilde{\mathcal{B}}_y|$ (dash), and for (right)
    $\mathrm{Re}\left\{\,\widetilde{\mathcal{B}}_x\,\right\}$ (solid)
    and $\mathrm{Im}\left\{\,\widetilde{\mathcal{B}}_y\,\right\}$
    (dash) for $S \sqrt{\varepsilon_0}\sin\theta_f \, t_e = 6$ and
    $\varepsilon_0 = 0.1$.  This case has $\theta_f=\theta_B=\pi/4$.}
  \label{fig:steady_distributions}
\end{figure}

Figure \ref{fig:steady_distributions} (left panel) shows the
distributions $D$ for the vector magnitude and for the directional
component magnitudes for a small value of $\varepsilon_0$.  These
distributions are smooth, positive, symmetric in $s$, and peak at
intermediate $\varepsilon/\varepsilon_0$.  $B^{rms}$ and the component
magnitudes are growing exponentially with time. We can fit this
with a cumulative growth rate, $\gamma^{rms} = t_e^{-1} \, \log
B^{rms}$, which we know from (\ref{eq:Gdistrib}) will scale as $S
\sqrt{\varepsilon_0} \sin\theta_f$.  For this value of $\varepsilon_0
= 0.1$, $|\widetilde{\mathcal{B}}_x|$ is smaller than
$|\widetilde{\mathcal{B}}_y|$, with a ensemble-mean ratio of 0.78.
For the leading eigenfunction in (\ref{eq:ef-te}), the anisotropy
ratio is
\begin{equation}
 \frac{|\widetilde{\mathcal{B}}_x|}{|\widetilde{\mathcal{B}}_y|} =
\frac{(1+\rmi s)E\sin\theta_f + \rmi \sqrt{2}\varepsilon 
\cos \theta_f \sin \theta_f}
{\sqrt{2}(1 - \rmi \varepsilon \cos^2\theta_f)} \,.
\end{equation}
For small $E$, the ratio tends to $E\sin\theta_f/\sqrt{2}$, which is
small; this is consistent with the anisotropy in Fig.
\ref{fig:mean-field}.  For large $E$, the ratio tends to
$|\tan\theta_f|$, which can have any value.  

What is the ensemble-mean magnetic field?  Its magnitude is
\begin{equation}
B^{mean} \, \equiv \, \Big| \, {\cal E}\Big[\, 
\widetilde{\vec{\mathcal{B}}}(t_e) \, \Big] \, \Big| \,
= \, \Big| \, \int_{-\infty}^{\infty} \, 
\widetilde{\vec{\mathcal{B}}}(\varepsilon,t_e) \, {\cal P}(\varepsilon) 
\,  | \, \rmd{}\varepsilon \,  \Big| \, \equiv \, \Big| \, 
\int_{-\infty}^{\infty} \, 
D[\,\widetilde{\vec{\mathcal{B}}} \,] \,  \rmd{}\varepsilon \, \Big| \,.
\label{eq:Bmean}
\end{equation}
Again this is evaluated with (\ref{eq:ef-te}).  We find that it too
exhibits exponential growth, so we fit a cumulative growth rate,
$\gamma^{mean}(t_e) = t_e^{-1} \, \log B^{mean} > 0$.  But the
ensemble mean growth is smaller than the ensemble r.m.s. growth, \ie
$\gamma^{mean} < \gamma^{rms}$.  The reason is illustrated in Fig.
\ref{fig:steady_distributions} (right panel) for the distributions of
two components,
$D[\,\mathrm{Re}\left\{\,\widetilde{\mathcal{B}}_x\,\right\}\,]$ and
$D[\,\mathrm{Im}\left\{\,\widetilde{\mathcal{B}}_y\,\right\}\,]$.
Their amplitude is comparable to the magnitude distributions in the
left panel, but they are oscillatory in $\varepsilon$ as a result of
$\cos[\gamma t_e]$ and $\sin[\gamma t_e]$ terms in (\ref{eq:ef-te}).
So the expected value from integration over $\varepsilon$ is small,
although not zero.  For Fig.  \ref{fig:steady_distributions},
$B^{mean} = 0.073 B^{rms}$, and $\gamma^{mean} = 0.76 \gamma^{rms}$.

\begin{figure}
  \includegraphics[width=0.49\textwidth]{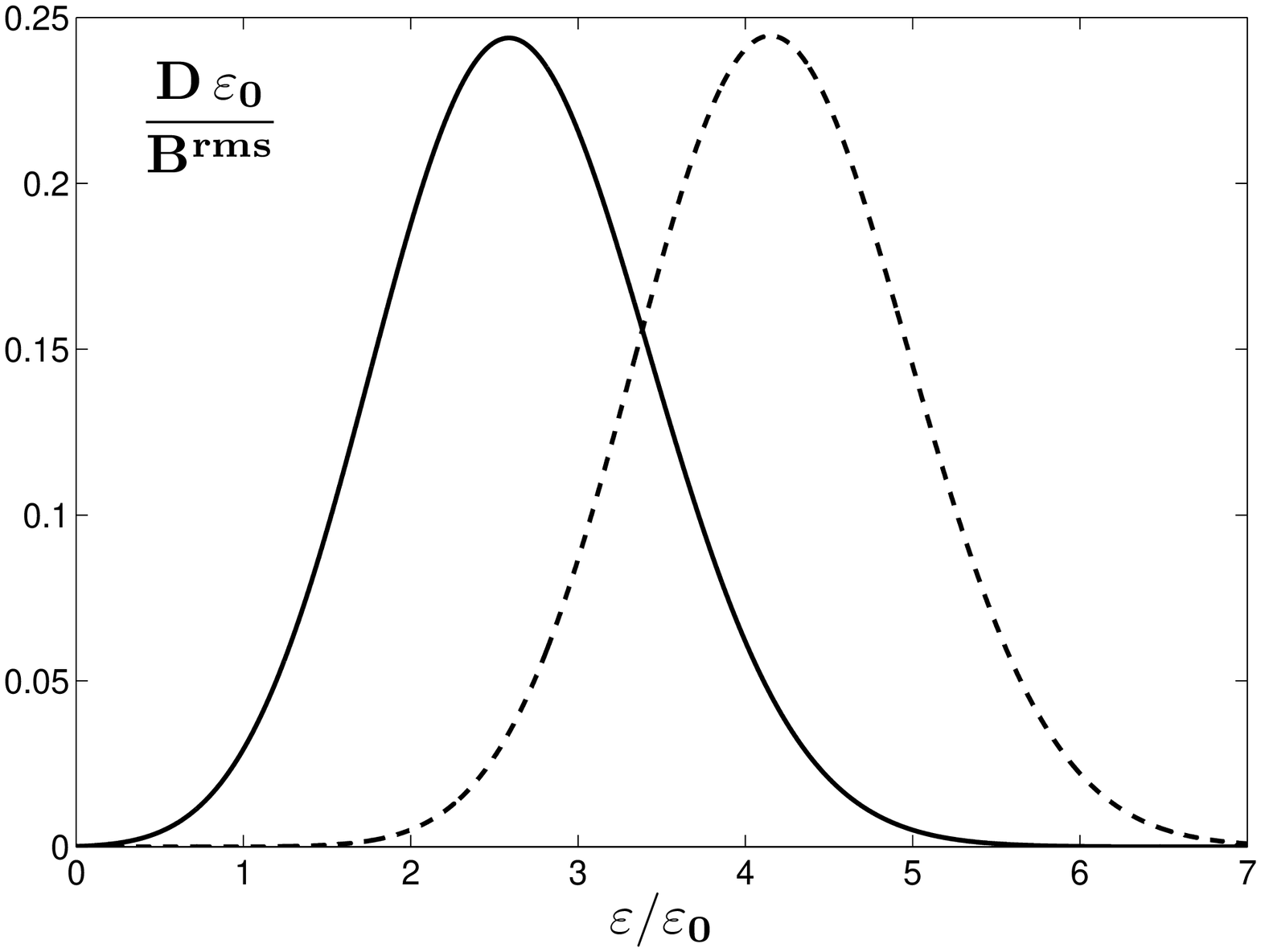}
  \hskip 1pt
  \includegraphics[width=0.49\textwidth]{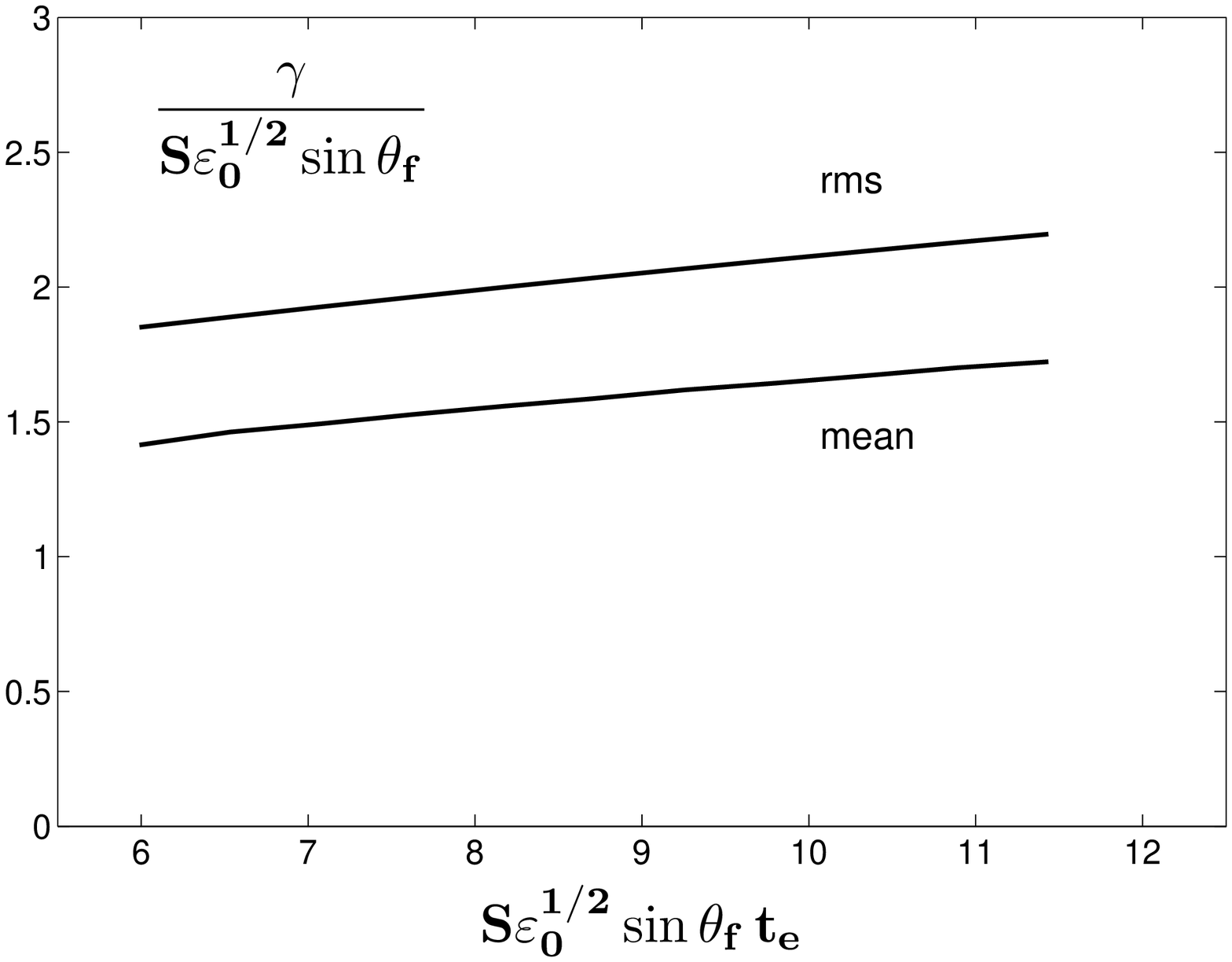}
  \caption{(Left) steady-forcing normalized distributions
    $D(\varepsilon)$ for $|\widetilde{\vec{\mathcal{B}}}|$ at two
    times: $S \sqrt{\epsilon_0}\sin\theta_f \, t_e = 6$ (solid) and
    $12$ (dash).  Case parameters are as in Fig.
    \ref{fig:steady_distributions}.  (Right) steady-forcing normalized
    growth rates, $\gamma^{rms}$ and $\gamma^{mean}$, as a function of
    evaluation time $t_e$. }
  \label{fig:steady_te}
\end{figure}

These relations are not sensitive to the initial condition
$\widetilde{\vec{\mathcal{B}}}(0)$, although it does influence the
partition among the real and imaginary parts of
$\widetilde{\vec{\mathcal{B}}}(t_e)$.  There are the expected
dependences of larger $\gamma$ with larger $S$ and $\varepsilon_0$ and
with $\theta_f$ closer to $\pi/2$, as in (\ref{eq:steady}).  With
larger $t_e$ the expected values are dominated by the farther tails of
the $D$ distributions, with slowly increasing $\gamma^{rms}(t_e)$ and
$\gamma^{mean}(t_e)$ associated with larger $\gamma(\epsilon)$ in the
tails (Fig. \ref{fig:steady_te}).  Even though larger $\epsilon$
values are less probable in $P(\varepsilon)$ in (\ref{eq:epspdf}),
they do have a more than compensating stronger dynamo growth rate that
emerges after long enough time.  Because the discrepancy between
$\gamma^{rms}$ and $\gamma^{mean}$ persists even in the
$P(\varepsilon)$ tail, the ratio $B^{mean}/B^{rms}$ decreases with
$t_e$ exponentially.  The steady-forcing dynamo does not become
independent of $t_e$ as $t_e \rightarrow \infty$, in contrast to the
finite-$t_f$ dynamo, in particular the small-$t_f$ dynamo analyzed in
Sec. \ref{sec:rapid}.

\subsubsection{Rapidly Varying Forcing} 
\label{sec:rapid}

The limiting forms for the ESD equation, (\ref{eq:complexB-eta}) and
(\ref{eq:complexB-eta-nu}), are also analyzable in the opposite limit
of $t_f \rightarrow 0$ by means of a cumulant expansion of a linear,
stochastic, ODE system \citep[Chap. XVI]{vanKampen07}.  For a
stochastic vector $\vec{A}(t)$ governed by
\begin{equation}
\pd_t \vec{A} = \left(\, \mathsfbi{L}_0 + \mathsfbi{L}_1(t) \,\right) \, 
	\cdot \vec{A} \,,
\label{eq:SDE}
\end{equation}
with the tensors $\mathsfbi{L}_0$ independent of time and $\mathsfbi{L}_1(t)$
a random stationary process with zero expected mean and finite
variance, the expected value ${\cal E}\Big[\vec{A}\Big]$ satisfies the
approximate deterministic ODE system,
\begin{equation}
\pd_t {\cal E}\Big[\vec{A}\Big] =  \left(\, \mathsfbi{L}_0 
+ \int_0^\infty \,  {\cal E}\Big[\mathsfbi{L}_1(t) \, 
\mathsfbi{L}_1(t-t') \Big] \, dt' 
+ \dots \,\right) \, \cdot {\cal E}\Big[\vec{A}\Big] \,,
\label{eq:vK}
\end{equation}
with the dots indicating neglected higher-order cumulant terms.  The
system (\ref{eq:vK}) has a time-independent matrix; hence, it has
eigenmodes with exponential time dependence with growth rates given by
the matrix eigenvalues.  The solution formula for ${\cal E}\Big[{\bf
  A}\Big](t)$ is called a time-ordered exponential matrix, and it has
a non-terminating series expansion with the leading terms as indicated
here.  The basis for the approximate neglect of the higher order terms
can be taken as the vanishing of ${\cal E}\Big[\mathsfbi{L}_11(t) \,
\mathsfbi{L}_1(t-t') \Big]$ except as $|t-t'| \rightarrow 0$.  In the
present situation with large $\nu$, this is equivalent to short
correlation times $t_f \rightarrow 0$ for the random forces,
$\hat{f}_z(t)$ and $\hat{o}_z(t)$, with $St_f \ll 1$ and $S/\nu \ll 1$
to be able to neglect higher-order products of $\mathsfbi{L}_0$ and
$\mathsfbi{L}_1$ in deriving (\ref{eq:vK}).

We apply (\ref{eq:SDE})-(\ref{eq:vK}) to (\ref{eq:complexB-eta-nu}) with
$\vec{A} = \widetilde{\vec{\mathcal{B}}}\,\exp\Bigl[k_z^2 {\cal
  F}_0^\dagger t /\eta\Bigr]$ with the following tensors:
\begin{equation}
\mathsfbi{L}_0 = S \, \begin{pmatrix} 0 & 0 \cr
             1 &  0 \cr \end{pmatrix} \,, \quad
\mathsfbi{L}_1 = \frac{2\rmi k_z{\cal H}^\dagger(t)}{\eta}  \, 
\begin{pmatrix} \cos\theta_f \sin\theta_f & \sin^2\theta_f \cr
 -\, \cos^2\theta_f & - \, \cos\theta_f \sin\theta_f \cr \end{pmatrix} \,.
\label{eq:L-B}
\end{equation}
This is a second-order, complex system.  We have made one {\it ad hoc}
simplification here, \viz replacing ${\cal F}^\dagger(t)$ by its
expected value, ${\cal F}_0^\dagger \equiv{\cal E}\Big[{\cal
F}^\dagger\Big] = 0.5$ from (\ref{eq:FHG_renorm}), and then factoring its
decay effect on $\widetilde{\vec{\mathcal{B}}}$ analogously to
(\ref{eq:factor-eta}).  The motivation is to simplify the analysis.
We already understand ${\cal F}$ as an eddy resistive damping.  This
role is played with qualitative fidelity by retaining only its mean
value, and anyway for large $\eta$ it is only a small increment to the
ordinary resistivity.  The result for (\ref{eq:vK}) is very simple
with (\ref{eq:L-B}) because $\mathsfbi{L}_1^{2} = 0$
independent of its time-variable prefactor, and
the eigenvalues of $\mathsfbi{L}_0$ are zero.  Hence, again after
restoring the resistive decay factors, the growth rate for ${\cal
E}\Big[\vec{\mathcal{B}}\Bigr]$ is
\begin{equation}
  \gamma = - \, \left(\eta + \frac{1}{2\eta} \right) \, 
	k_z^2 \le 0 \,;
\end{equation}
\ie in this ($\eta \rightarrow \infty$, $t_f \rightarrow 0$) limit there 
is only resistive decay of the expected value of the mean magnetic
field, weakly augmented by the eddy resistive effect.

We could continue the cumulant expansion for $\vec{\mathcal{B}}$ and
(\ref{eq:SDE}) to higher orders in $St_f$ and $S/\nu$
\citep{vanKampen07}, seeking growth in the ensemble-mean,
large-scale field, ${\cal E}\Big[\vec{\mathcal{B}}\Bigr]$, but its
$\gamma$ would be small in these parameters compared to the growth in
the mean magnetic variance, ${\cal E}\Big[\, |\vec{\mathcal{B}}|^2 \,
\Bigr]$.  To obtain a dynamo result for the latter, we instead apply
(\ref{eq:SDE})-(\ref{eq:vK}) to the fourth-order real covariance
system derived from (\ref{eq:complexB-eta}) for the vector,
\begin{equation}
\vec{A} = \left(\ |\widetilde{B}_x|^2 , \ |\widetilde{B}_y|^2 , \
\mathrm{Re}[\widetilde{B}_x^\ast\widetilde{B}_y], \
\mathrm{Im}[\widetilde{B}_x^\ast\widetilde{B}_y] \ \right) \, \times \,
\exp\Bigl[2 k_z^2 {\cal F}_0^\dagger t/\eta\Bigr] \,,
\end{equation}
again factoring out the mean eddy resistive effect with the
simplification ${\cal F}^\dagger(t) \approx {\cal F}^\dagger_0 = 0.5$.
The associated tensors are defined by
\begin{eqnarray}
& \mathsfbi{L}_0 = S \mathsfbi{L}_0^\dagger , \,  
\mathsfbi{L}_0^\dagger = \begin{pmatrix} 0 & 0 & 0 & 0 \cr
                                  0 & 0 & 2 & 0 \cr
                                  1 & 0 & 0 & 0 \cr 
                                  0 & 0 & 0 & 0 \cr \end{pmatrix} \,,
	\nonumber \\ 
& \mathsfbi{L}_1 = 
\frac{-2k_z{\cal H}^\dagger(t)}{\eta} \mathsfbi{L}_1^\dagger , 
\,  \mathsfbi{L}_1^\dagger =
   \begin{pmatrix} 0 & 0 & 0 & 2\sin^2\theta_f \cr
                   0 & 0 & 0 & 2\cos^2\theta_f \cr
                   0 & 0 & 0 & - 2\cos\theta_f \sin\theta_f \cr
 \cos^2\theta_f & \sin^2\theta_f & 2\cos\theta_f \sin\theta_f & 0 \cr
   \end{pmatrix} .
\label{eq:Lcov}
\end{eqnarray}
The expectation value in (\ref{eq:vK}) applied to $\mathsfbi{L}_1(t)
\mathsfbi{L}_1(t-t')$ acts entirely on its scalar prefactor in 
(\ref{eq:Lcov}) because its matrix factor $\mathsfbi{L}_1^\dagger$ is
deterministic and time-independent.  We evaluate the corresponding
scalar prefactor that arises in (\ref{eq:vK}) as
\[
\frac{4k_z^2}{\eta^2} \, \int_0^\infty \, {\cal E}\Big[{\cal H}^\dagger(t)
{\cal H}^\dagger(t-t')\Bigr] \, dt' \,.
\]
Tracing backwards through the forcing relations (\ref{eq:helicity}),
(\ref{eq:f-renorm}), and (\ref{eq:FHG_renorm}), we derive
\begin{equation}
{\cal E}\Big[{\cal H}^\dagger(t) {\cal H}^\dagger(t-t')\Bigr] =  
0.5 \, {\cal E}\Bigl[|\hat{f}_z^\dagger|^2\Bigr] \, 
{\cal E}\Bigl[|\hat{o}_z^\dagger|^2\Bigr] \, 
\exp\Bigl[-2|t'|/t_f\Bigr] \,,
\end{equation}
utilizing the fact that the real and imaginary parts of $\hat{f}_z$
and $\hat{o}_z$ are independent, stationary processes each with an
exponential correlation time $t_f$ as in (\ref{eq:fcor}).  After
performing the time integration with this expression, the value of the
preceding prefactor is
\begin{equation}
\frac{2k_z^2\,t_f}{\eta^2} \, {\cal E}\Bigl[{\cal H}^{\dagger 2}\Bigr]  
= \frac{k_z^2\,t_f}{\eta^2}\,,
\label{eq:prefactor}
\end{equation}
because ${\cal E}\Bigl[{\cal H}^{\dagger 2}\Bigr] = 0.5 \, F_z\dagger
\, O_z^\dagger = 0.5$ from (\ref{eq:f-renorm}).  This completes the
specification of the deterministic, time-independent matrix in
(\ref{eq:vK}) for the covariance system as
\begin{equation}
\mathsfbi{L} = S \mathsfbi{L}_0^\dagger + \frac{k_z^2\,t_f}{\eta^2}
\mathsfbi{L}_1^{\dagger 2} \,.
\label{eq:vKmatrix}
\end{equation}
We evaluate its eigenvalues $\Gamma$ analytically from
$det[\mathsfbi{L} - \Gamma \mathsfbi{I}] = 0$, which is a
fourth-order polynomial equation.  We can factor a $\Gamma = 0$ root,
leaving a third-order system with the reduced form of $\Gamma^3 + p
\Gamma = q$ for coefficients $p \propto S$ and $q \propto S^2$.  
With a simplification provided by the prefactor (\ref{eq:prefactor})
being small compared to $S$, we can neglect the $p$ term and obtain
the approximate solution,
\begin{equation}
\Gamma \approx q^{1/3} = \left(\, 
\frac{2 k_z^2 S^2 \sin^4\theta_f \, t_f}{\eta^2} \right)^{1/3} \,.
\label{eq:Gamma}
\end{equation}
This approximation is consistent with finite $S$, small $t_f$ and
$k_z$, and large $\eta$; recall that we also assume $St_f, \ S/\nu
\ll 1$ for the leading order cumulant approximation (\ref{eq:vK}).  The 
three solutions (\ref{eq:Gamma}) are one with real, positive $\Gamma$ (\ie
a dynamo) and a complex conjugate pair with $\mathrm{Re}[\Gamma] <
0$.  We divide the positive eigenvalue $\Gamma$ by 2 and restore the
resistive decay factors to obtain the growth rate for the r.m.s. value
of the mean field, $\left( \, {\cal
E}\Big[|\vec{\mathcal{B}}|^2\Bigr]\,\right)^{1/2}$:
\begin{equation}
  \gamma = - \, \left(\eta + \frac{1}{2\eta} \right) k_z^2  
\ + \ \left(\, \frac{k_z^2S^2 
\, \sin^4\theta_f \, t_f }{4\eta^2} \,  \right)^{1/3} \,.
\label{eq:smalltc}
\end{equation}
A dynamo can occur with $\gamma > 0$ if there are both forcing
helicity and shear and if $k_z$ is small but nonzero; this behavior is
the same as in the steady-forcing dynamo (\ref{eq:steady}) for this
same limiting ESD system (\ref{eq:complexB-eta-nu}), as well as for
the general dynamo in Sec. \ref{sec:general}.  In this limit of small
correlation time with zero mean helicity and finite helicity variance,
the expected value for the mean field $\vec{\mathcal{B}}$ does not
grow, but the expected value for the mean magnetic energy
$\vec{\mathcal{B}}^2$ does.  The steady-forcing dynamo also has a much
smaller ensemble mean than r.m.s. (Sec. \ref{sec:steady}).

Besides the leading eigenvalue (\ref{eq:Gamma}), we can obtain the
associated eigenfunction for the matrix (\ref{eq:vKmatrix}).  With the
same approximation of a small prefactor for $\mathsfbi{L}^{\dagger
  2}$, we derive the following for the expected ratio of component
variances,
\begin{align}
{\cal E}\Bigl[ |\widetilde{B}_x|^2 \Bigr] & \approx \frac{2 \sin^4\theta_f}{\Gamma}
\ {\cal E}\Bigl[ |\widetilde{B}_y|^2 \Bigr]
    \nonumber \\
& = \left(\frac{2k_z^2 \sin^4\theta_f t_f}{S\eta^2}\right)^{2/3} 
\,  {\cal E}\Bigl[ |\widetilde{B}_y|^2 \Bigr] \,.
\label{eq:Bxyratio}
\end{align}
Thus, the streamwise mean magnetic energy is small compared to the
transverse energy in the present limit with transient forcing, small
$k_z$ and $t_f$, and large $\eta$.  The small ratio is also consistent
with the previous example of dynamo behavior with more general
parameters in Fig.  \ref{fig:mean-field}, as well as with the
steady-forcing dynamo in Sec. \ref{sec:steady} when $\varepsilon_0$ is
small.

As with the steady forcing (\ref{eq:steady-opt}) we can optimize the
growth rate in $k_z$:
\begin{align}
k_{z \, opt} &= \left(\, \frac{S^2 \sin^4\theta_f \, t_f }
{108 \, \eta^2 \, (\eta + \frac{1}{2\eta})^3} \, \right)^{1/4}
	\nonumber \\
\gamma_{opt} &= \left(\,
\frac{S^2 \sin^4\theta_f \, t_f }
{27\, \eta^2(\eta+ \frac{1}{2\eta})} \, \right)^{1/2} \,.
\label{eq:short-opt}
\end{align}
The parameter tendencies here all have the same signs as with
steady-forcing and with the general ESD (Sec. \ref{sec:general}), but
the exponents are different in the two $t_f$ limits.  In particular,
the optimal growth rate dependences are 
\begin{align}
\gamma &\sim \ S \ \ || {\cal
H} || \, \eta^{-3/2} \, k_{yf}^2 \, t_f^{1/2} \qquad \quad {\rm as} 
\ t_f \rightarrow 0 
	\nonumber \\
\gamma &\sim \ S^{2/3} \, || {\cal H}^\dagger ||^{2/3} \,
\eta^{-1} \, k_{yf}^{4/3} \, t_f^0 \qquad {\rm as} 
\ t_f \rightarrow \infty \,,
\end{align}
where the norm symbol $|| \cdot ||$ denotes the r.m.s. or mean
magnitude as appropriate, and we have formally restored the helicity
variance factor $|| {\cal H} ||$ for emphasis.  In both cases the
growth rate $\gamma$ is vanishingly small as $\eta \rightarrow
\infty$, $S \rightarrow 0$, $|| {\cal H} || \rightarrow 0$, or
$\theta_f \rightarrow 0, \pi$, and for the short correlation time
case, $\gamma$ is small as $t_f \rightarrow 0$.  For non-limiting
values of the parameters, however, $\gamma$ is not small
(Sec. \ref{sec:general}).  We reiterate that there is no dependence of
$\gamma$ on $\nu$ in the limit $\nu \rightarrow \infty$, independent
of the value of $t_f$.

As with the steady forcing limit, an analogy exists between the
fluctuating helicity ESD in (\ref{eq:complexB-eta-nu}) and a low-order
ODE fluctuating alpha--omega dynamo {\it ansatz}
\citep{Vishniac97,Silantev00} (also called the incoherent alpha--shear
dynamo).  Therefore, from a historical perspective of astrophysical
dynamo theory, we see that the ESD in (\ref{eq:complexB}) provides
both a theoretical justification for the alpha--omega {\it ansatz},
with an explicit characterization of the relevant shearing-wave
velocity fluctuations, and a generalization to finite Reynolds numbers
(\ie $\eta,\nu < \infty$).

In summary, these two different $t_f$ limits with analytic dynamo
solutions for the large-$(\eta,\nu)$ ESD (\ref{eq:complexB-eta-nu})
show qualitatively similar but functionally different parameter
tendencies in $S$, $\eta$, $k_z$, and $\theta_f$; anisotropy with
$|\widetilde{\mathcal{B}}_y|$ usually larger than
$|\widetilde{\mathcal{B}}_x|$; and an ensemble-mean magnetic energy,
${\cal E}\Bigl[\, |\widetilde{\vec{\mathcal{B}}}|^2 \,\Bigr]$, much
larger than the energy of the ensemble-mean field, $|\, {\cal E}\Bigl[
|\widetilde{\vec{\mathcal{B}}} \Bigr]\,|^2$.  These characteristics
carry over to the more general ESD solutions in Sec.
\ref{sec:general}.

\subsection{Other Limit Pathways}
\label{sec:limit.S}

The preceding ESD derivation of (\ref{eq:complexB2}) assumes $k_{yf}L
\gg 1$ to assure $\aver{exp[\rmi(\phi+\phi')]}^{x,y} \approx 0$ and
$k_{yf}LS \ {\rm min}[t_f,\ 1/\nu] \gg 1$ to assure
$\aver{exp[\rmi(\phi+\phi')]}^{x,y} \ne 0$ for selected time arguments
of the phases $\phi(\mu)$ and $\phi(\mu')$.  The latter assumption
yields (\ref{eq:x-avg}), which is useful in simplifying the
normalization condition (\ref{eq:general_renorm}) for $KE$ and
compacting the ESD equation (\ref{eq:complexB2}) for
$\widetilde{\vec{\mathcal{B}}}$ by reducing the number of time history
integrals in the mean electromotive force curl (Appendix
\ref{sec:appA}).  We prefer the physical rationale of this pathway
based only on a primary assumption of large $L$, consistent with
uniform mean shear and no boundary conditions, because it does not
constrain the values of the other parameters that are physically more
meaningful than $L$.  The result is independent of $L$ itself.  The
further ESD simplifications in Sec. \ref{sec:limit.L} follow from
$\eta,\nu \rightarrow \infty$.

However, this is not a unique pathway for deriving ESD equations that
are essentially similar.  In particular, neither of the limits $S
\rightarrow 0$ nor $t_f \rightarrow 0$ is problematic even though they
appear inconsistent with the second assumption above.  As previously
explained, we do require $\nu > 0$ for statistical equilibration of
velocity fluctuations and $k_{yf} \ne 0$ for nontrivial shear tilting
and dynamo behavior.

Shear tilting makes $\phi(t)$ in (\ref{eq:phi-cons}) or
(\ref{eq:phase}) a continuous function of time.  When $S=0$,
$\phi=\phi_f$, and the average of the differenced-phase factor is
$\aver{exp[\rmi(\phi-\phi')]}^{x,y} = 1$ for all time arguments.  When
$S \rightarrow 0$ as a primary assumption, this relation is
approximately true.  We still require the weaker assumption about
large domain size, $k_{yf}L \gg 1$, to be able to neglect the
summed-phase factors, $\aver{exp[\rmi(\phi+\phi')]}^{x,y}$.  Even with
these phase averaging relations resolved, further assumptions are
needed to compact the electromotive forcing, and large $\eta$ and/or
$\nu$ suffice.  The outcome is equivalent to
(\ref{eq:complexB-eta-nu}) with dynamo solutions when $S>0$.  If
instead the primary assumption is $t_f \rightarrow 0$ in
combination with $k_{yf}L \gg 1$, then the requirement on the average
of the differenced-phase factor in the $KE$ normalization is resolved
with an approximate integral over the forcing correlation factor,
$\exp[-|\mu-\mu'|/t_f]$, in (\ref{eq:general_renorm}), but this
assumption is not enough to compact the electromotive force curl.
Again this can be accomplished with additional assumptions of large
$\eta$ and/or $\nu$, leading to the equivalents of
(\ref{eq:complexB-eta}) with shear tilting and
(\ref{eq:complexB-eta-nu}) without it.  In neither of these limits is
there a compact equivalent to the general ESD (\ref{eq:complexB2})
with finite $\eta$ and $\nu$.  Also, because the dynamo solutions of
(\ref{eq:complexB-eta-nu}) have $\gamma$ small with $S$ and $t_f$,
this derivation pathway is not as physically germane as the primary
one in Sec. \ref{sec:limit.L}.

Yet another derivation pathway assumes finite $L$ and spatially
periodic boundary conditions in shearing coordinates with discretized
shearing-frame wavenumbers with $\Delta k = 2\pi/L$.  If the forcing is
at one of the discretized wavenumbers at least in $k_{yf}$, then the
spatial average of the summed-phase factor vanishes.  To accommodate
continuous shear tilting in the finite Fourier series representation,
the forcing amplitude time series is viewed as impulses at discrete
times, $t_m = t_0 + m\Delta t$, $\Delta t = 2\pi/S k_{yf}L$,
$m=0,1,2,\dots $, when a discrete shearing-frame $x$-wavenumber
$k_{xm} = k_{xf}+ Sk_{yf}t_m$ (or its periodic alias) coincides with
$k_{xf}$ in the laboratory frame.  (This discretization is the one
used in a MHD computational code with a finite number of Fourier modes
\citep{Yousef08b}.)  This allows the shearing-coordinate spatial
average of the differenced-phase factors to have the requisite
property for a compact ESD derivation.  The resulting ESD replaces the
time-history integrals with finite sums over $m$ at discrete forcing
times $t_m$, and it replaces the continuous laboratory-frame
$\vec{k}(t-\mu)$ with $\vec{k}(t-t_m)$.  This pathway retains the
familiar dependence on $L$ for a discrete Fourier series; this
dependence disappears as $L \rightarrow \infty$ when the
shearing-periodicity pathway merges with the large-domain pathway as
$\Delta k$ and $\Delta t$ vanish.  The general behaviors of the
finite-$L$ shearing-periodicity ESD and $L \rightarrow \infty$ ESD in
(\ref{eq:complexB2}) are essentially the same.  Because of the
simplicity of the spatial averaging with the shearing-periodic
boundary conditions and the analytical advantages of the assumptions
of large $\eta$ and $\nu$, small $S \ne 0$, and small $t_f$, a
proof-of-concept ESD exposition is in \citet{Heinemann11a}.  Its
solution coincides with Sec. \ref{sec:rapid}.  Notice that this
combined pathway achieves spatial homogeneity even without
the enlarged ensemble of uniform mean flows in $\vec{V}$
(Sec. \ref{sec:KE_ND}).

\section{General Parameter Dependences}
\label{sec:general}

With the normalization conditions
(\ref{eq:normalize})-(\ref{eq:part-norm}), the non-dimensional
parameters of the ESD equation (\ref{eq:complexB2}) are $S$, $\nu$,
$t_f$, $\theta_f$, $k_z$, $\eta$, and $\theta_B$.  {\it A priori} we
are interested in possible dynamo behavior over their full ranges.
Section \ref{sec:dynamo} shows a typical ``mid-range'' example by
computational integration, and Sec. \ref{sec:limit} has analytic
formulas for the parameter dependences of the growth rate $\gamma$ in
two asymptotic limits associated with $\eta, \nu \rightarrow \infty$
and $t_f \rightarrow 0$ or $\infty$.  In this section we survey the
parameters space computationally to show that $\gamma$ in the ESD
solution is a smooth, simple function of all its parameters.

For given parameters, a computational solution provides a particular
realization of the random forcing in Sec. \ref{sec:force}.  When there
is exponential growth in $|\vec{\mathcal{B}}(t)|$, a fit $\propto
e^\gamma t$ is made over a long integration period (\eg $S \Delta t =
10^3$ in Fig. \ref{fig:mean-field}).  The $\gamma$ value varies from
one realization to another, but the results we report here are fairly
well determined, as indicated by the smoothness of parameter curves
based on separate estimations at separate parameters.  Nevertheless,
it is computationally laborious to obtain an ensemble perspective over
many realizations.

\begin{figure}
  \centerline{\includegraphics[width=0.7\textwidth]{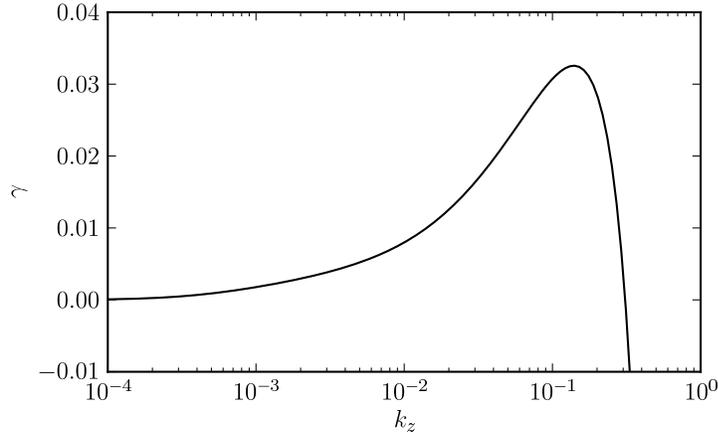}}
  \caption{Mean-field growth rate $\gamma$ as a function of $k_z$.
  Parameter values are $S=1$, $\theta_f=\pi/4$, $t_f=0.1$, $\nu=0.1$,
  and $\eta=0.1$.}  
\label{fig:gamma-kz}
\end{figure}

Dynamo growth occurs for finite values of $0< k_z < k_f = 1$
(Fig. \ref{fig:gamma-kz}); \ie increasing $k_z$ amplifies the
fluctuating helical forcing in (\ref{eq:complexB2}) that is essential
to the ESD, and dynamo growth is quenched by resistive decay when
$k_z$ is too large.  There is an optimal intermediate value for $k_z$
where $\gamma$ is a maximum.  This behavior is approximately the same
as evident in the analytic solutions in Secs. \ref{sec:steady} and
\ref{sec:rapid}.

\begin{figure}
  \centerline{\includegraphics[width=0.7\textwidth]{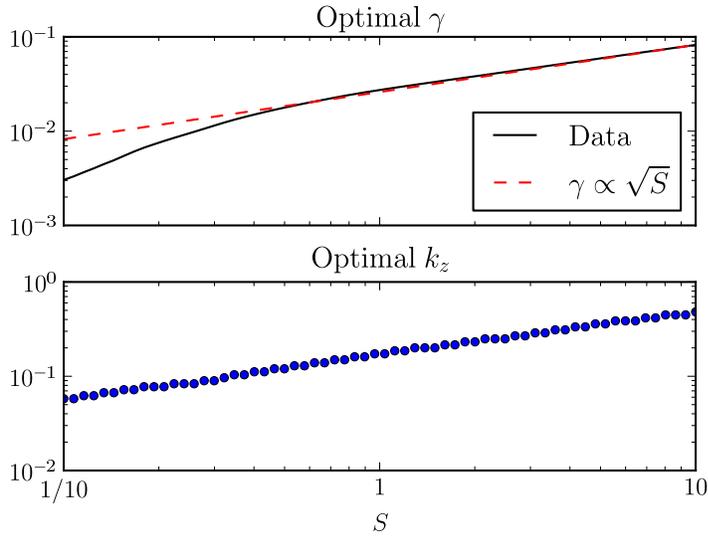}}
  \caption{Mean-field growth rate $\gamma$ associated with the optimal
  $k_z$ value as a function of shear $S$.  Other parameters are
  $\theta_f=\pi/4$, $t_f=0.1$, $\nu=0.1$, and $\eta=0.1$.  
  The dots here (and in subsequent figures) indicate the sampling
  density for this evaluation of the ESD.}  
\label{fig:gamma-S}
\end{figure}

The functional dependence of $\gamma$ on the shear $S$ is in Fig.
\ref{fig:gamma-S}, based on optimization over $k_z$ with the other
parameters held fixed.  The dynamo growth rate increases monotonically
with $S$; the slope of $\gamma(S)$ decreases for larger $S$.  A
power-law fit to $\gamma(S)$ shows an exponent approximately in the
range 0.5--1, which is consistent with the values of 2/3 and 1 in
the limiting formulas (\ref{eq:steady-opt}) and (\ref{eq:short-opt})

The associated optimal $k_z(S)$ is always small relative to $k_f=1$,
and it too increases with $S$.  A power law fit shows an exponent
similar to the limit values of 1/2 and 1/3 in (\ref{eq:steady-opt})
and (\ref{eq:short-opt}).  In the ESD there is no threshold in $S$ for
dynamo growth, given sufficiently small $k_z \ne 0$.  With either
$S=0$ or $k_z =0$, there is no dynamo.  $\gamma(k_z)$ is a convex
function of $k_z$ that vanishes when $k_z$ is not small as well as
when $k_z \rightarrow 0$; this is a similar shape as in the limit
formulas (\ref{eq:steady}) and (\ref{eq:smalltc}).  For all other
parameters held fixed (including $k_z$), there is a minimum threshold
value of $S$ for dynamo action, as is also true in the limit formulas
(\ref{eq:steady}) and (\ref{eq:smalltc}).

\begin{figure}
  \centerline{\includegraphics[width=0.7\textwidth]{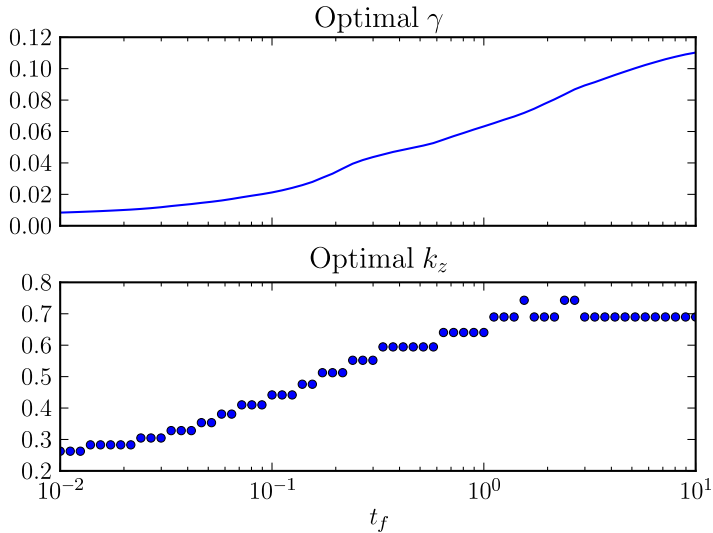}}
  \caption{Mean-field growth rate $\gamma$ associated with the optimal
    $k_z$ value as a function of forcing correlation time $t_f$.
    Other parameters are $\theta_f=\pi/4$, $S=2$, $\nu=10$,
    $\eta=0.01$, and $\Delta t=0.01$.}
  \label{fig:gamma-tau}
\end{figure}

The dependence of $\gamma$ on the forcing correlation time $t_f$ is in
Fig.  \ref{fig:gamma-tau}, again based on optimization over $k_z$.
$\gamma$ and $k_z$ both increase with $t_f$.  This tendency is
consistent at small $t_f$ with the limit formulas in
(\ref{eq:short-opt}).  For larger $t_f$ values the slope of
$\gamma(t_f)$ increases with $t_f$ in the range surveyed here,
although we know from (\ref{eq:steady-opt}) that $\gamma$ asymptotes
to a finite value with steady forcing.  The optimal $k_z(t_f)$ levels
off with large $t_f$, here at a value only slightly smaller than $k_f =
1$; this behavior is not anticipated by the limit formulas in
Sec. \ref{sec:limit} that indicate small $k_z$ for large $\eta$.

We demonstrate the roles of the forcing components $\hat{f}_z$ and
$\hat{o}_z$ by alternately setting them to zero.  $\hat{o}_z = 0$
removes all forcing from (\ref{eq:complexB2}), hence has no effect on
$\vec{\mathcal{B}}$.  $\hat{f}_z = 0$ retains the forcing in ${\cal F}$
but makes ${\cal H} = 0$; in this case $\vec{\mathcal{B}}(t)$ shows
algebraic growth in time but no dynamo. Thus, a dynamo requires both
$u_z$ and $\vec{u}_\perp$ to be nonzero.  By keeping both components
non-zero but arbitrarily setting ${\cal F} = 0$ with ${\cal H} \ne 0$
in (\ref{eq:complexB2}), $\gamma$ is modestly increased; this confirms
the interpretation of the ${\cal F}$ effect as turbulent resistivity
that weakens dynamo growth (Sec. \ref{sec:limit}).  If ${\cal F}(t)$ is
replaced by its time-mean value, the dynamo behavior is essentially
the same.

\begin{figure}
  \centerline{\includegraphics[width=0.7\textwidth]{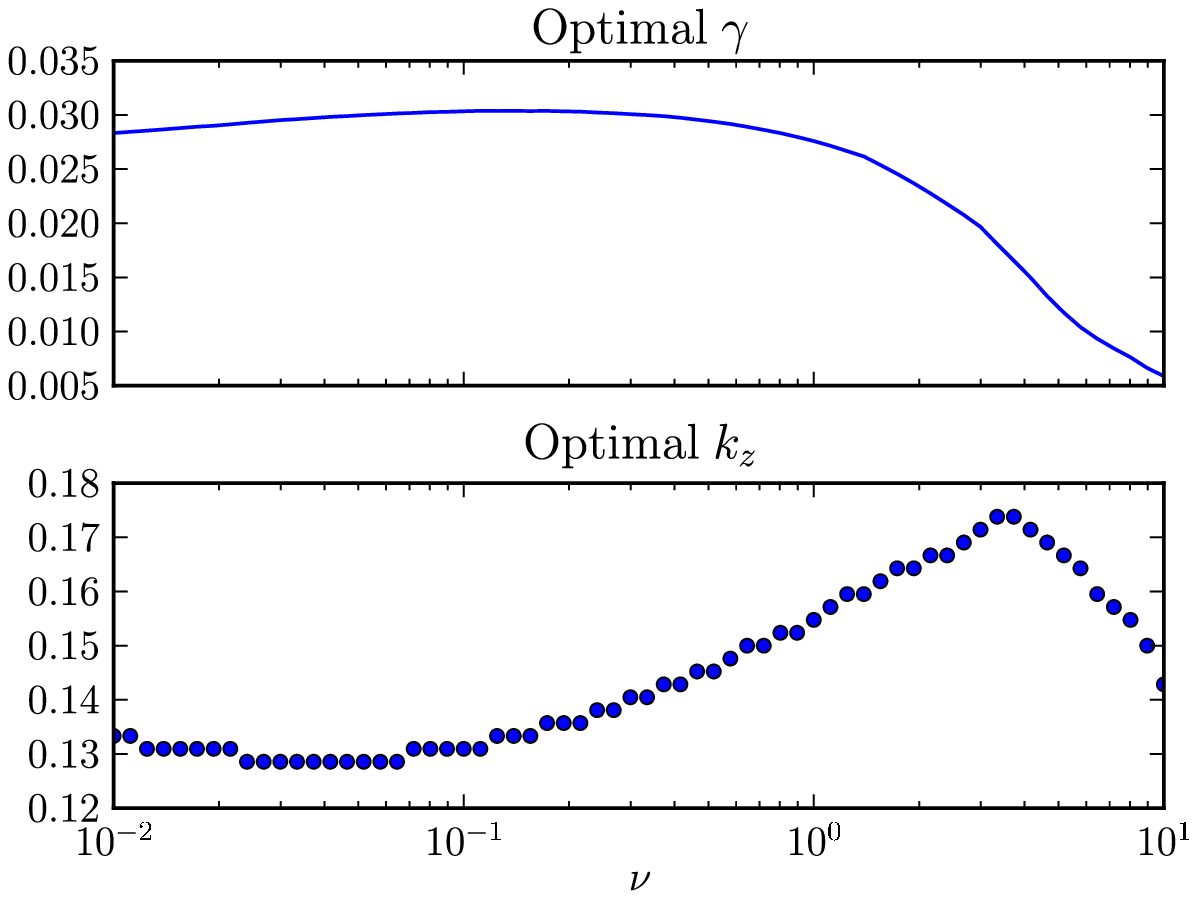}}
  \caption{Mean-field growth rate $\gamma$ associated with the optimal
    $k_z$ value as a function of viscosity $\nu$.  Other parameters
    are $\theta_f=\pi/4$, $S=1$, $t_f=0.1$, $\eta=0.1$, and $\Delta
    t=0.025$.  }
  \label{fig:gamma-nu}
\end{figure}

Viscous and resistive diffusion both diminish dynamo growth, but they
do not suppress it entirely (Figs.
\ref{fig:gamma-nu}-\ref{fig:gamma-eta}).  The growth rate becomes
independent of $\nu \rightarrow 0$ for fixed $\eta$, and it becomes
independent of $\eta \rightarrow 0$ for fixed $\nu$.  The latter
indicates that the ESD is a ``fast'' dynamo with $\gamma \ne 0$ as
$\eta \rightarrow 0$ (Roberts and Soward, 1992).  At the other
extreme, to sustain a dynamo as $\eta \rightarrow \infty$, the value
of $k_z(\eta)$ must become very small so that resistive decay is not
dominant; this is consistent with the limit formulas
(\ref{eq:steady-opt}) and (\ref{eq:short-opt}), where $\gamma(\eta)$
decreases as a power law with exponents of -1 and -5/2, respectively.
$\gamma(\nu)$ decreases with $\nu$ for large $\nu$.  We can take the
$\nu \rightarrow \infty$ limit of (\ref{eq:complexB2}) for general
$\eta$, using the same type of approximation procedure as at the
beginning of Sec.  \ref{sec:limit}.  The key approximation
in this limit is
\begin{equation}
\int^t_0 \, \rmd{}\lambda \, \int^\lambda_0 \rmd{}\mu \, 
	G_\nu^\dagger(\lambda-\mu)G_\nu^\dagger(t-\mu) 
\rightarrow  \frac{1}{\nu} \,,
\end{equation}
with $\mu, \lambda \rightarrow t$ for the arguments of the other
integrand factors.  The resulting $(L,\nu)$-limit mean-field equation
has the same structure as (\ref{eq:complexB-eta-nu}) except now the
electromotive force curl has a prefactor of $1/\nu$ instead of
$1/\eta$.  Consequently, $\gamma(\nu)$ must decrease with large $\nu$
as in Fig.  \ref{fig:gamma-nu}.

The optimal $\gamma(\theta_f)$ and $k_z(\theta_f)$ are both largest
for intermediate $\theta_f$ values (Fig. \ref{fig:gamma-theta}).  The
limit formulas predict a peak at $\theta_f = \pi/2$ and $\gamma = 0$
at $\theta_f = 0, \pi$ ($k_{yf} = 0$).  However, these limits are
based on (\ref{eq:complexB-eta-nu}) after $\nu \rightarrow \infty$,
which suppresses any effect of shear tilting in the ESD.  In the more
general case an up-shear orientation ($0 < \theta_f \pi/2$) is more
conducive to dynamo growth.  Thus, the Orr effect of phase tilting in
shearing waves (Sec. \ref{sec:con-wave}) augments the dynamo
efficiency.  This is because, when $\theta_f$ is up-shear, the helical
forcing factor transiently increases in magnitude as $k_x(t)$
decreases between $t = 0$ and $t = 1/\tan\theta_f/S>0$ when $k_x(t)$
passes through zero and thereafter becomes increasingly large and
negative.  This has the effect of transiently augmenting the effective
helicity, hence dynamo forcing, compared to a down-shear case where
$|k_x(t)|$ monotonically increases and the effective helicity only
decreases with time.  The magnitude of this transient dynamo
enhancement is limited by the viscous decay that ensues during the
phase tilting toward $k_x = 0$ (and beyond), consistent with the Orr
effect disappearing when $\nu \rightarrow \infty$.

From an ensemble of numerical integrations, we find that the estimate
mean value of $\gamma$ is independent of $\theta_B$; \ie the initial
conditions of $\vec{\mathcal{B}}$ are not important for the dynamo
apart from the necessity of a seed amplitude in $\vec{\mathcal{B}}$ to
enable the dynamo.

\begin{figure}
  \centerline{\includegraphics[width=0.8\textwidth]{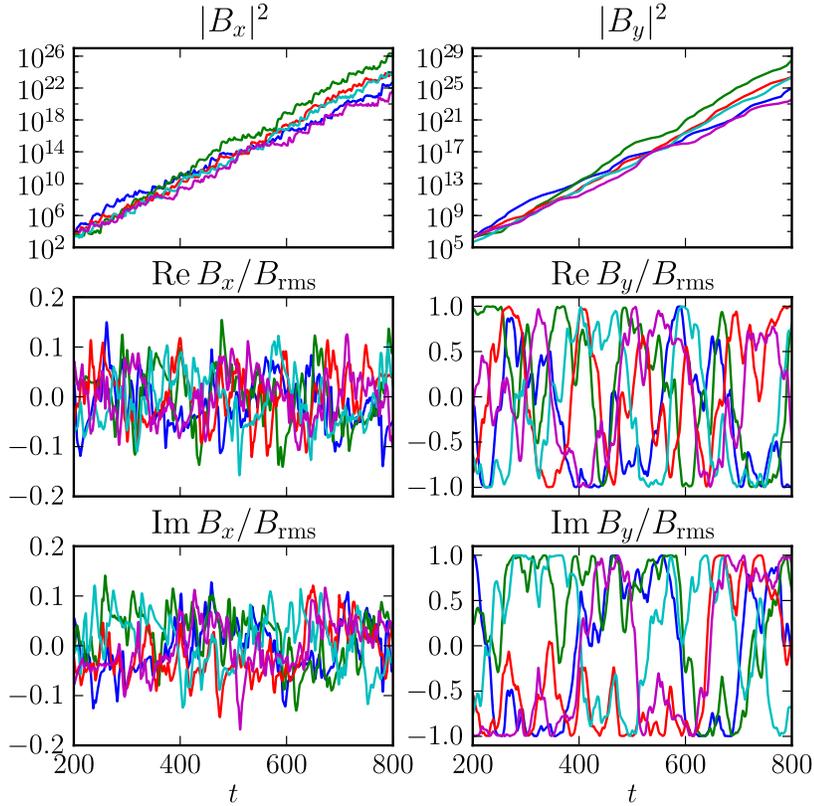}}
  \caption{Five ESD realizations in different colors of the mean field
  component variances, $|\mathcal{B}_x|^2(t)$ and
  $|\mathcal{B}_y|^2(t)$ (top row), and of their real and imaginary
  parts normalized by $B_{rms} = |\vec{\mathcal{B}}|(t)$ (bottom two
  rows).  Parameters are $k_z = 0.14$, $S=1$, $\theta_f=\pi/4$,
  $t_f=0.1$, $\nu=0.1$, and $\eta=0.1$.  All realizations have the same
  initial condition $\vec{\mathcal{B}}(0)$.} 
\label{fig:ensemble}
\end{figure}

The analytic solutions in Sec. \ref{sec:limit.L} for the $\eta,\nu
\rightarrow \infty$ limit show that the ensemble mean field, ${\cal
E}\Big[\,\vec{\mathcal{B}}\,\Big]$, has a smaller (but nonzero) dynamo
growth rate $\gamma$ than the r.m.s. field for a steady-forcing
ensemble as well as a smaller (but undetermined) $\gamma$ for
rapidly-varying forcing.  Figure \ref{fig:ensemble} illustrates, for a
more generic parameter set, how the components of the complex
amplitude $\vec{\mathcal{B}}(t)$ vary substantially both with time and
among different realizations, including spontaneous sign reversals on
a time scale longer than those directly related to the parameters (\ie
the non-dimensional fluctuation turn-over time of 1, as well as $t_f$,
$1/S$, $1/\eta$, and $1/\nu$); long-interval reversals also occur for
Earth's magnetic field.  This occurs even as the mean magnetic field
amplitude inexorably grows, albeit with evident but relatively modest
low-frequency and inter-realization variability.  It has proved to be
computationally difficult to accurately determine the ensemble mean of
$\vec{\mathcal{B}}$ over many random realizations for fixed initial
conditions in the general ESD (\ref{eq:complexB2}).  Our computational
experience is consistent with the mean field magnitude typically being
only a small fraction of the square root of the mean magnetic energy.
Thus, the ESD with random small-scale forcing is essentially a random
large-scale dynamo.

\section{Summary and Prospects}
\label{sec:summary}

We derive the Elemental Shear Dynamo (ESD) model for a random
barotropic force with a single horizontal wavevector in a steady flow
with uniform shear in a large domain.  It is a quasi-linear theory
that is rigorously justified for vanishing magnetic Reynolds number
($1/\eta \rightarrow 0$) and experimentally supported for more general
parameters.  It robustly exhibits kinematic dynamo behavior as long as
the force $\vec{f}$ has both vertical and horizontal components with
finite forcing helicity variance; the vertical wavenumber $k_z$ of the
initial seed amplitude of the mean magnetic field
$\aver{\vec{B}}^{x,y}$ is nonzero but small compared to the horizontal
wavenumber of the forcing; and the forcing wavenumber orientation is
not shear-normal (\ie $k_{yf} \ne 0$).  When these conditions are
satisfied, the dynamo growth rate is larger when $S$ is larger, the
resistivity $\eta$ and viscosity $\nu$ are smaller, the forcing
correlation time $t_f$ is larger, and the forcing wavenumber
$\theta_f$ is in an upshear direction.  The ensemble-mean of the
energy of the horizontally averaged magnetic field grows as a dynamo,
but the energy of the ensemble-mean magnetic field is much smaller.
Reversals in $\aver{\vec{B}}^{x,y}(t)$ are common over time intervals
long compared to $t_f$.  Because the growth-rate curves have broad
maxima in both parameters and fluctuation wavenumbers
(Sec. \ref{sec:general}), we expect dynamo action with a broad
spectrum in $\vec{k}_f$ and $k_z$, consistent with the quasi-linear
superposition principle (\ref{eq:superpose}).

The ESD ingredients of small-scale velocity fluctuations and
large-scale shear are generic across the universe, so its dynamo
process is likely to be relevant to the widespread existence of
large-scale magnetic fields.  Of course, the simple spatial symmetries
assumed in the ESD model are a strong idealization of natural flows,
and the ESD is not a general MHD model because of its quasi-linearity
assumptions.  Investigation of more complex situations is needed to
determine the realm of relevance for the ESD behavior shown here,
especially in turbulent flows with intrinsic variability and large
Reynolds number.

{\bf Acknowledgments:} This work benefited greatly from
extensive discussions with Tobias Heinemann, who also helped with some
of the calculations and figures, and with Alexander Schekochihin, who
has led our inquiry into the shear dynamo. I also appreciate a long
and fruitful partnership with Steven Cowley on dynamo behaviors, first
at small scales and now at large.  This paper is a fruit of
unsponsored research.

\appendix
\section{Derivation of $\vec{F}_\mathcal{B}$ in (\ref{eq:FBevaluation})}
\label{sec:appA}

This appendix fills in steps between the formal expression for the
curl of the mean electromotive force (\ref{eq:FBdef}) and its
particular expression in the ESD (\ref{eq:FBevaluation}).  Here we
retain the convention that all vectors are horizontal.

To provide a more compact notation, we rewrite the vertical phase
factor coefficient for the fluctuation field (\ref{eq:mag-fluc-2}) as
\begin{align}
\vec{b}(x,y,t) &= 
	\int^t_0 \, \rmd{}\lambda \, \int^\lambda_0 \rmd{}\mu \
\Bigl[\ \vec{b}_{+}(t,\lambda,\mu) 
\, e^{\rmi \phi(\mu)}
+ \ \vec{b}_{-}(t,\lambda,\mu) 
\, e^{- \rmi \phi(\mu)} \ \Bigr] \,,
\end{align}
where
\begin{align}
\vec{b}_{+} &=\frac{1}{2} \, G_\eta(t-\mu,\lambda-\mu)G_\nu(\lambda-\mu) \
 	\Bigl[\ k_z \Bigl(\, - \, \rmi \hat{f}_z(\mu) 
	\mathsfbi{S}(t-\lambda) \cdot \vec{\mathcal{B}}(\lambda) \,\Bigr)
\nonumber \\
&\ + \, \frac{\vec{e}_z \times \vec{k}(t-\mu)}{k^2(\lambda-\mu)} 
\, \Bigl(\, \hat{o}_z(\mu) \, 
	( \vec{k}(\lambda-\mu) \cdot \vec{\mathcal{B}}(\lambda) )
 \,\Bigr) \ \Bigr]
	\nonumber \\
&
	\nonumber \\
\vec{b}_{-} &= \frac{1}{2} \, G_\eta(t-\mu,\lambda-\mu)G_\nu(\lambda-\mu) \
 	\Bigl[\ k_z \Bigl(\, - \, \rmi \hat{f}_z^\ast(\mu) 
	\mathsfbi{S}(t-\lambda) \cdot \vec{\mathcal{B}}(\lambda) \,\Bigr)
\nonumber \\
&\ + \, \frac{\vec{e}_z \times \vec{k}(t-\mu)}{k^2(\lambda-\mu)} 
\, \Bigl(\, \hat{o}_z^\ast(\mu) \, 
	( \vec{k}(\lambda-\mu) \cdot \vec{\mathcal{B}}(\lambda) )
 \,\Bigr) \ \Bigr] \,.
\label{eq:bpm-def}
\end{align}

We evaluate the three terms for $\aver{\, \vec{F}_\mathcal{B}
  \,}^{x,y}$ in (\ref{eq:FBdef}) for each of the terms in
$\vec{b}_{+}$ and $\vec{b}_{-}$.  To do so involves the spatial
average of products of factors with exponential phase functions, $\rmi
( \, \pm \, \phi + k_z z)$.  Employing (\ref{eq:x-avg}) we will make
use of the general identities,
\begin{align}
& \aver{\, 
\mathrm{Re}\left\{ A_1(\rho) \, e^{\rmi \phi(\rho)}\right\}
\ \mathrm{Re}\left\{ A_2(\mu) \, e^{\rmi \phi(\mu)}
e^{\rmi k_z z}\right\} 
\,}^{x,y} 
= \frac{C_L}{2}\, \delta(\rho-\mu) \, 
\mathrm{Re}\left\{ A_1^\ast A_2
e^{\rmi k_z z}\right\}
	\nonumber \\
&
	\nonumber \\
&\aver{\, \mathrm{Re}\left\{ A_1(\rho) \, e^{\rmi \phi(\rho)}\right\}
\ \mathrm{Re}\left\{ A_2(\mu) \, e^{-\rmi \phi(\mu)}
e^{\rmi k_z z}\right\} \,}^{x,y} 
= \frac{C_L}{2}\, \delta(\rho-\mu) \, 
\mathrm{Re}\left\{ A_1 A_2
e^{\rmi k_z z}\right\} \,.
\label{eq:ReRe}
\end{align}

The first term in (\ref{eq:FBdef}) is evaluated as follows:
\begin{align}
& - \, \la (\vec{u} \cdot{\nabla})\vec{b}^\prime \ra^{x,y} 
	\nonumber \\
& \quad = - \, \la \left(\, \int_0^t \,\rmd{}\rho \, G_\nu(t-\rho) \,
\left(\frac{- \, \vec{e}_z \times \vec{k}(t-\rho)}
	{k^2(t-\rho)}\right) \, \mathrm{Re}\left\{\,\rmi
	\hat{o}_z(\rho) \, e^{\rmi \phi(\rho) } \,\right\} 
\cdot \vec{k}(t-\mu) \,\right) \ 
	\nonumber \\
&\qquad \qquad \qquad \int^t_0 \, \rmd{}\lambda \, \int^\lambda_0 \rmd{}\mu \
\Bigl[\ \mathrm{Re}\left\{\rmi\vec{b}_{+}(t,\lambda,\mu) 
\, e^{\rmi \phi(\mu)}e^{\rmi k_z z}\right\}
	\nonumber \\
&\qquad \qquad \qquad \qquad + \ \mathrm{Re}\left\{
-\rmi\vec{b}_{-}(t,\lambda,\mu) \, e^{- \rmi \phi(\mu)}
e^{\rmi k_z z}\right\} \ \Bigr]
\ra^{x,y}
	\nonumber \\
& \quad = \ \frac{C_L}{2} \, 
\int^t_0 \, \rmd{}\lambda \, \int^\lambda_0 \rmd{}\mu \, G_\nu(t-\mu) \,
\left(\frac{\vec{e}_z \times \vec{k}(t-\mu) \cdot \vec{k}(t-\mu)}
	{k^2(t-\rho)}\right) 
	\nonumber \\
& \qquad \qquad \qquad \qquad \mathrm{Re}\left\{\, 
\left( \hat{o}_z^\ast \vec{b}_{+} +  \hat{o}_z 
\vec{b}_{-}\right) \, e^{\rmi k_z z}\,\right\}  
	\nonumber \\
& \quad = \ 0 \,.
\end{align}
The formula (\ref{eq:ReRe}) is used to obtain the middle right-hand
side, and the final result comes from the identify, $\vec{e}_z \times
\vec{a} \cdot \vec{a} = 0$.

The second term in (\ref{eq:FBdef}) is evaluated as follows:
\begin{align}
& - \, \la (u_z \pd_z )\vec{b}^\prime \ra^{x,y} 
	\nonumber \\
& \quad = - \, \la \left(\, \int_0^t \,\rmd{}\rho \, G_\nu(t-\rho) \,
\mathrm{Re}\left\{\, \hat{f}_z(\rho) \, e^{\rmi \phi(\rho) } \,\right\} 
\, k_z \,\right) \ 
	\nonumber \\
&\qquad \qquad \int^t_0 \, \rmd{}\lambda \, \int^\lambda_0 \rmd{}\mu \
\Bigl[\ \mathrm{Re}\left\{\rmi \vec{b}_{+}(t,\lambda,\mu) 
\, e^{\rmi \phi(\mu)}e^{\rmi k_z z}\right\}
	\nonumber \\
&\qquad \qquad \qquad \qquad + \ \mathrm{Re}\left\{\rmi 
\vec{b}_{-}(t,\lambda,\mu) \, e^{- \rmi \phi(\mu)}
e^{\rmi k_z z}\right\} \ \Bigr]
\ra^{x,y}
	\nonumber \\
& \quad = - \, \frac{C_L}{2} \, 
\int^t_0 \, \rmd{}\lambda \, \int^\lambda_0 \rmd{}\mu \, G_\nu(t-\mu) k_z \,
\mathrm{Re}\left\{\, \rmi
\left( \hat{f}_z^\ast \vec{b}_{+} +  \hat{f}_z \vec{b}_{-}\right) \, 
e^{\rmi k_z z}\,\right\}
	\nonumber \\
& \quad = - \, \frac{C_L}{2} \, 
\int^t_0 \, \rmd{}\lambda \, \int^\lambda_0 \rmd{}\mu \,
	G_\eta(t-\mu,\lambda-\mu)G_\nu(\lambda-\mu)G_\nu(t-\mu) \,
	\nonumber \\
& \qquad \qquad\qquad
\Bigl[\ |\hat{f}_z|^2(\mu)\, k_z^2  \,
   \mathrm{Re}\left\{\mathsfbi{S}(t-\lambda) \cdot \vec{\mathcal{B}}(\lambda)
	e^{\rmi k_z z}\right\}
	\nonumber \\
& \ + \,
 \mathrm{Re}\left\{\hat{f}_z^\ast(\mu)\hat{o}_z(\mu)\right\} \, 
  k_z \, \vec{e}_z \times \vec{k}(t-\mu)
\, \Bigl(\, 
\frac{\vec{k}(\lambda-\mu)}{k^2(\lambda-\mu)} \cdot \mathrm{Re}
\left\{\rmi \vec{\mathcal{B}}(\lambda)e^{\rmi k_z z}\right\}
\,\Bigr) \ \Bigr] \,.
\end{align}
The formula (\ref{eq:ReRe}) is used to obtain the second right-hand
side, and (\ref{eq:bpm-def}) is substituted to obtain the final result,
which agrees with the first and second terms in (\ref{eq:FBevaluation}).

The final term in (\ref{eq:FBdef}) is evaluated as follows:
\begin{align}
& \la (\vec{b}^\prime \cdot{\nabla})\vec{u} \ra^{x,y} 
	\nonumber \\
& \quad = \, \la \int^t_0 \, \rmd{}\lambda \, \int^\lambda_0 \rmd{}\mu \,
\int_0^t \,\rmd{}\rho \, G_\nu(t-\rho) \,
\left(\frac{- \, \vec{e}_z \times \vec{k}(t-\rho)}
	{k^2(t-\rho)}\right) \, \mathrm{Re}\left\{\,\rmi\, \rmi \,
	\hat{o}_z(\rho) \, e^{\rmi \phi(\rho) } \,\right\}
	\nonumber \\
&\qquad \qquad \qquad \Bigl(\, \vec{k}(t-\rho) \ \cdot \
\Bigl[\ \mathrm{Re}\left\{\vec{b}_{+}(t,\lambda,\mu) 
\, e^{\rmi \phi(\mu)}e^{\rmi k_z z}\right\}
	\nonumber \\
&\qquad \qquad \qquad \qquad \qquad \qquad  
+ \ \mathrm{Re}\left\{
\vec{b}_{-}(t,\lambda,\mu) \, e^{- \rmi \phi(\mu)}
e^{\rmi k_z z}\right\} \ 
\Bigr] \,\Bigr) \ra^{x,y}
	\nonumber \\
& \quad = \  \frac{C_L}{2} \,
\int^t_0 \, \rmd{}\lambda \, \int^\lambda_0 \rmd{}\mu \, G_\nu(t-\mu) \,
\frac{\vec{e}_z \times \vec{k}(t-\mu)}
	{k^2(t-\mu)} \,
	\nonumber \\
& \qquad \qquad \qquad \Bigl(\, \vec{k}(t-\mu) \ \cdot \
\mathrm{Re}\left\{\, \left( \hat{o}_z^\ast\vec{b}_{+} + 
\hat{o}_z\vec{b}_{-} \right)\, e^{\rmi k_z z}\,\right\}
\, \Bigr)
	\nonumber \\
& \quad = - \, \frac{C_L}{2} \,
\int^t_0 \, \rmd{}\lambda \, \int^\lambda_0 \rmd{}\mu \, 
G_\eta(t-\mu,\lambda-\mu)G_\nu(\lambda-\mu)G_\nu(t-\mu) \,
\frac{\vec{e}_z \times \vec{k}(t-\mu)}
	{k^2(t-\mu)} \,
	\nonumber \\
& \qquad \qquad \qquad
k_z \mathrm{Re}\left\{ \hat{f}_z^\ast \hat{o}_z \right\} \,
\Bigl(\, \vec{k}(t-\mu) \ \cdot \mathrm{Re}\left\{\, \rmi
\mathsfbi{S}(t-\lambda) \cdot \vec{\mathcal{B}}(\lambda) 
e^{\rmi k_z z}\,\right\} \,\Bigr) \,.
\end{align}
Again, (\ref{eq:ReRe}) is used to obtain the second right-hand side,
and (\ref{eq:bpm-def}) is substituted to obtain the final result,
which agrees with the third term in (\ref{eq:FBevaluation}).
In this substitution, the terms in $\vec{b}_\pm \ \propto \
\hat{o}_z, \ \hat{o}_z^\ast$ do not survive because
they yield a factor, $\vec{e}_z \times \vec{k}(t-\mu)
\cdot \vec{k}(t-\mu) = 0$.

This completes the derivation of (\ref{eq:FBevaluation}).

\section{Computational Solution of (\ref{eq:complexB2})}
\label{sec:method}

The ESD solutions in Sec. \ref{sec:dynamo} are obtained by numerical
integration of the integro-differential equation system
(\ref{eq:complexB2}).  This system is potentially expensive to solve
because of the two time integrals, requiring $O(T^3)$ operations to
integrate to time $T$.  We convert this to an $O(T)$ system (formally
comparable to the size an ODE integration, albeit with a much larger
coefficient for $T$) by limiting the integration range to fixed
intervals, $t-\tau \le \lambda, \le t$ and $t-\tau \le \mu \le
\lambda$, once $t>\tau$; for smaller $t$ values, the integrations
start from $\lambda = \mu = 0$.  A sufficient motivation for this
approximation is that the two viscous decay factors (\ref{eq:Enudef})
become vanishingly small for large values of its arguments $t-\mu$ and
$\lambda-\mu$ in (\ref{eq:complexB2}).  For a given $\nu$ value, we
determine $\tau$ by the requirement that $G_\nu(\tau)
\le\hat{\epsilon}\ll1$. In practice we typically choose
$\hat{\epsilon}=10^{-7}$ and make sure that the results do not change
significantly if we further decrease the value of $\hat{\epsilon}$.

The domain of integration is a triangle in $(\lambda,\mu)$-space.
Because of this, the forcing functions $\mathcal{F}(\mu)$ and
$\mathcal{H}(\mu)$ only need to be retained in memory for the range
$t-\tau\le\mu\le{}t$ to evaluate
$\widetilde{\vec{\dot{\mathcal{B}}}}(t)$ and advance
$\widetilde{\vec{\mathcal{B}}}(t)$ in time.  With the restricted
integration intervals, the mean field equation (\ref{eq:complexB2}) is
\begin{equation}
  \label{eq:mean-field-PQ}
  \vec{\dot{\widetilde{\mathcal{B}}}}(t) =
  S \widetilde{\mathcal{B}}_x(t)\, \vec{e}_y
  + \int_{t-\tau}^t\!\rmd\lambda\int_{t-\tau}^\lambda\!\rmd\mu\,
  \Bigl[
      \mathcal{F}(\mu)\mathcal{P}(t-\mu,\lambda-\mu) +
  \rmi\mathcal{H}(\mu)\mathcal{Q}(t-\mu,\lambda-\mu)
  \Bigr]
  \cdot\widetilde{\vec{\mathcal{B}}}(\lambda) \,,
\end{equation}
where we have introduced the second-order matrices,
\begin{equation}
  \label{eq:P-matrix}
  \mathcal{P}(t_1,t_2) =
  -k_z^2 G_\eta(t_1,t_2) G_\nu(t_1) G_\nu(t_2)
  \mathcal{S}(t_1-t_2)
\end{equation}
and
\begin{equation}
  \label{eq:Q-matrix}
  \mathcal{Q}(t_1,t_2) =
  -k_z G_\eta(t_1,t_2) G_\nu(t_1) G_\nu(t_2)
  \Bigl[k^{-2}(t_1) + k^{-2}(t_2)\Bigr]
  (\vec{e}_z\times\vec{k}(t_1)) \, \vec{k}(t_2) \,.
\end{equation}
We may convert the double time integral in (\ref{eq:mean-field-PQ}) to a
double time integral `into the past' via the substitutions
$\lambda'=t-\lambda$ and $\mu'=t-\mu$, giving
\begin{equation}
  \label{eq:mean-field-past}
  \vec{\dot{\widetilde{\mathcal{B}}}}(t) =
  S \widetilde{\mathcal{B}}_x(t)\vec{e}_y
  + \int_0^\tau\!\rmd\lambda'\int_{\lambda'}^\tau\!\rmd\mu'\,
  \Bigl[
      \mathcal{F}(t-\mu')\mathcal{P}(\mu',\mu'-\lambda') +
  \rmi\mathcal{H}(t-\mu')\mathcal{Q}(\mu',\mu'-\lambda')
  \Bigr]
  \cdot\widetilde{\vec{\mathcal{B}}}(t-\lambda') \,.
\end{equation}
Note that, because $0\le\lambda'\le\mu'\le\tau$, the matrices (\ref{eq:P-matrix})
and (\ref{eq:Q-matrix}) can be evaluated once and for all in the ranges
$0\le{}t\le\tau$ and $0\le\lambda\le\tau$ at the beginning of the simulation.

To discretize (\ref{eq:mean-field-past}) in time, we write this
equation as a system of one integro-differential and one integral equation,
\viz
\begin{equation}
  \label{eq:mean-field-B}
  \vec{\dot{\widetilde{\mathcal{B}}}}(t) =
  S \widetilde{\mathcal{B}}_x(t)\vec{e}_y
  - \int_0^\tau\!\rmd\lambda\,
  \mathcal{G}(t,\lambda)\cdot\widetilde{\vec{\mathcal{B}}}(t-\lambda)
\end{equation}
and
\begin{equation}
  \label{eq:mean-field-G}
  \mathcal{G}(t,\lambda) =
  \int_\lambda^\tau\!\rmd\mu
  \Bigl[
  \mathcal{F}(t-\mu)\mathcal{P}(\mu,\mu-\lambda) +
  \rmi\mathcal{H}(t-\mu)\mathcal{Q}(\mu,\mu-\lambda)
  \Bigr] \,,
\end{equation}
where we have now dropped the primes from $\lambda$ and $\mu$. Using the
trapezoidal rule, a second-order accurate representation of
(\ref{eq:mean-field-B}) is given by
\begin{multline}
  \frac
  {\vec{\mathcal{B}}^{n+1} - \mathcal{S}(\Delta t)\cdot\vec{\mathcal{B}}^{n}}
  {\Delta t} =
  \mathcal{S}(\Delta t)\cdot\left(
  \frac{\Delta t}{4}
  \mathcal{G}^{n,0}\cdot\vec{\mathcal{B}}^n +
  \frac{\Delta t}{2}
  \sum_{m=1}^{K-1}\mathcal{G}^{n,m}\cdot\vec{\mathcal{B}}^{n-m}
  \right)
  \\
  +\left(
  \frac{\Delta t}{4}
  \mathcal{G}^{n+1,0}\cdot\vec{\mathcal{B}}^{n+1} +
  \frac{\Delta t}{2}
  \sum_{m=1}^{K-1}\mathcal{G}^{n+1,m}\cdot\vec{\mathcal{B}}^{n+1-m}
  \right) \,,
\label{eq:B7}
\end{multline}
where
$\widetilde{\vec{\mathcal{B}}}^n=\widetilde{\vec{\mathcal{B}}}(n\Delta{}t)$
and $\mathcal{G}^{n,m}=\mathcal{G}(n\Delta{}t,m\Delta{}t)$; we
have anticipated (\ref{eq:G-nK}) below. The integer $K$ is defined
through the relation $\tau=K\Delta{}t$. The matrix factor
$\mathcal{S}(\Delta{}t)$ arises from treating the shear stretching
term exactly. Because (\ref{eq:B7}) is linear, it may be easily
solved for $\widetilde{\vec{\mathcal{B}}}^{n+1}$ provided the matrix
$\mathcal{G}^{n+1,0}$ can be inverted. To compute
$\mathcal{G}^{n,m}$, we again use the trapezoidal rule to obtain
\begin{subequations}
\begin{align}
  \mathcal{G}^{n,m} &=
  \frac{\Delta t}{2}\left(
  \mathcal{F}^{n-m}\mathcal{P}^{m,0} + \rmi\mathcal{H}^{n-m}\mathcal{Q}^{m,0}
  \right) \nonumber\\ &+
  \Delta t\sum_{l=m+1}^{K-1}\left(
  \mathcal{F}^{n-l}\mathcal{P}^{l,l-m} + \rmi\mathcal{H}^{n-l}\mathcal{Q}^{l,l-m}
  \right) \nonumber\\ &+
  \frac{\Delta t}{2}\left(
  \mathcal{F}^{n-K}\mathcal{P}^{K,K-m} + \rmi\mathcal{H}^{n-K}\mathcal{Q}^{K,K-m}
  \right) \\
  \label{eq:G-nK}
  \mathcal{G}^{n,K} &= 0 \,.
\end{align}
\end{subequations}

\bibliographystyle{jfm}
\bibliography{esd_references}

\begin{thebibliography}{14}
\expandafter\ifx\csname natexlab\endcsname\relax\def\natexlab#1{#1}\fi

\bibitem[Brandenburg \& Subramanian(2005)]{Brandenburg05}
{\sc Brandenburg, A. \& Subramanian, K.} 2005 Astrophysical magnetic fields and
  nonlinear dynamo theory. {\em Physics Reports\/} {\bf 417}, 1--209.

\bibitem[Heinemann {\em et~al.\/}(2011{\natexlab{{\em a\/}}})Heinemann,
  McWilliams \& Schekochihin]{Heinemann11a}
{\sc Heinemann, T., McWilliams, J.~C. \& Schekochihin, A.} 2011{\natexlab{{\em
  a\/}}} Magnetic field generation by randomly forced shearing waves. Submitted
  to Phys. Rev. Lett.

\bibitem[Heinemann {\em et~al.\/}(2011{\natexlab{{\em b\/}}})Heinemann,
  McWilliams \& Schekochihin]{Heinemann11b}
{\sc Heinemann, T., McWilliams, J.~C. \& Schekochihin, A.} 2011{\natexlab{{\em
  b\/}}} The shear dynamo in $2^{+}$ dimensions. In preparation.

\bibitem[Krause \& Radler(1980)]{Krause80}
{\sc Krause, F. \& Radler, K.~H.} 1980 {\em Mean-Field Magnetohydrodynamics and
  Dynamo Theory\/}. Pergamon Press.

\bibitem[Kulsrud(2010)]{Kulsrud10}
{\sc Kulsrud, R.~H.} 2010 The origin of our galatic magnetic field. {\em
  Astron. Nachr.\/} {\bf 331}, 22--26.

\bibitem[Moffatt(1978)]{Moffatt78}
{\sc Moffatt, H.~K.} 1978 {\em Magnetic Field Generation in Electrically
  Conducting Fluids\/}. Cambridge University Press.

\bibitem[Parker(1971)]{Parker71}
{\sc Parker, E.~N.} 1971 The generation of magnetic fields in astrophysical
  bodies. {II}. {T}he galactic field. {\em Ap. J.\/} {\bf 163}, 255--278.

\bibitem[Roberts \& Soward(1992)]{Roberts92}
{\sc Roberts, P.~H. \& Soward, A.~M.} 1992 Dynamo theory. {\em Ann. Rev. Fluid
  Mech.\/} {\bf 24}, 459--512.

\bibitem[Silant'ev(2000)]{Silantev00}
{\sc Silant'ev, N.A.} 2000 Magnetic dyanmo due to turbulent helicity
  fluctuations. {\em Astron. Astrophys.\/} {\bf 364}, 339--347.

\bibitem[Sridhar \& Subramanian(2009)]{Sridhar09}
{\sc Sridhar, S. \& Subramanian, K.} 2009 Nonperturbative quasi-linear approach
  to the shear dynamo problem. {\em Phys. Rev. E\/} {\bf 80}, 0066315.

\bibitem[{van Kampen}(2007)]{vanKampen07}
{\sc {van Kampen}, N.} 2007 {\em Stochastic Processes in Physics and
  Chemistry\/}. Elsevier, 3rd edition.

\bibitem[Vishniac \& Brandenburg(1997)]{Vishniac97}
{\sc Vishniac, E. \& Brandenburg, A.} 1997 An incoherent $\alpha-\omega$ dynamo
  in accretion disks. {\em Ap. J.\/} {\bf 475}, 263--274.

\bibitem[Yousef {\em et~al.\/}(2008{\natexlab{{\em a\/}}})Yousef, Heinemann,
  Schekochinin, Kleeorin, Rogachevskii, Cowley \& McWilliams]{Yousef08b}
{\sc Yousef, T.~A., Heinemann, T., Schekochinin, A.~A., Kleeorin, N.,
  Rogachevskii, I., Cowley, S.~C. \& McWilliams, J.~C.} 2008{\natexlab{{\em
  a\/}}} Numerical experiments on dynamo action in sheared and rotating
  turbulence. {\em Astron. Nachr.\/} {\bf 329}, 737--749.

\bibitem[Yousef {\em et~al.\/}(2008{\natexlab{{\em b\/}}})Yousef, Heinemann,
  Schekochinin, Kleeorin, Rogachevskii, Iskakov, Cowley \&
  McWilliams]{Yousef08a}
{\sc Yousef, T.~A., Heinemann, T., Schekochinin, A.~A., Kleeorin, N.,
  Rogachevskii, I., Iskakov, A.~B., Cowley, S.~C. \& McWilliams, J.~C.}
  2008{\natexlab{{\em b\/}}} Generation of magnetic field by combined action of
  turbulence and shear. {\em Phys. Rev. Lett.\/} {\bf 100}, 184501.

\end{thebibliography}


\end{document}